\documentclass[a4paper,11pt]{article}
\pdfoutput=1 

\usepackage{jheppub} 
\usepackage[T1]{fontenc} 
\usepackage{subcaption}
\usepackage{mathrsfs}
\graphicspath{ {graphs/} }

\preprint{PUPT-2575}

\title{\boldmath \LARGE Recursion Relations in Witten Diagrams and Conformal Partial Waves}

\author[]{Xinan Zhou}

\affiliation[]{Princeton Center for Theoretical Science, Princeton University, \\Princeton, NJ 08544, U.S.A.}
\affiliation[]{C. N. Yang Institute for Theoretical Physics, Stony Brook University, \\Stony Brook, NY 11794, U.S.A.}

\emailAdd{xinanz@princeton.edu}

\abstract{We revisit the problem of performing conformal block decomposition of exchange Witten diagrams in the crossed channel. Using properties of conformal blocks and Witten diagrams, we discover infinitely many linear relations among the crossed channel decomposition coefficients. These relations allow us to formulate a recursive algorithm that solves the decomposition coefficients in terms of certain seed coefficients. In one dimensional CFTs, the seed coefficient is the decomposition coefficient of the double-trace operator with the lowest conformal dimension. In higher dimensions, the seed coefficients are the coefficients of the double-trace operators with the minimal conformal twist.  We also discuss the conformal block decomposition of a generic contact Witten diagram with any number of derivatives. As a byproduct of our analysis, we  obtain a similar recursive algorithm for decomposing conformal partial waves in the crossed channel.

 }

\begin{document}
\maketitle
\flushbottom
\section{Introduction and Summary}
Witten diagrams are the AdS analogue of position space Feynman diagrams in flat space, and appear as the building blocks in the holographic computation of conformal correlators in the supergravity limit. These diagrams were extensively studied  in the early days of AdS/CFT in position space, but their simplicity was only later appreciated in Mellin space \cite{Mack:2009mi,Penedones:2010ue,Paulos:2011ie,Fitzpatrick:2011ia}. Recently there has been another resurgence of interest in the community to revisit the Witten diagrams.
 The renewed interest is partly due to the curious appearance of these diagrams in the Mellin bootstrap method \cite{Gopakumar:2016wkt,Gopakumar:2016cpb,Gopakumar:2018xqi}, where the tree-level exchange Witten diagrams (modulo certain ambiguities in adding contact terms) are used as an expansion basis for conformal correlators. Relatedly, the usual conformal blocks admit a natural AdS bulk description in terms of a variation of the exchange Witten diagrams. They are the so-called geodesic Witten diagrams where the integration regions of the cubic vertex points are restricted to the AdS geodesics connecting the boundary operators \cite{Hijano:2015zsa}.\footnote{See also \cite{Rastelli:2017ecj,Goncalves:2018fwx} for generalization of geodesic Witten diagrams to CFTs with conformal boundaries and defects.}  Moreover, one is also motivated to revisit these diagrams because of the remarkable simplicity recently discovered in the  holographic one-half BPS four-point functions \cite{Rastelli:2016nze,Alday:2017xua,Aprile:2017bgs,Aprile:2017xsp,Rastelli:2017udc,Alday:2017vkk,Aprile:2017qoy,Rastelli:2017ymc,Zhou:2017zaw,Zhou:2018ofp,Aprile:2018efk,Caron-Huot:2018kta,Alday:2018pdi}.\footnote{Modern methods have been recently invented to efficiently compute  holographic correlators at the tree level. See \cite{Rastelli:2016nze,Rastelli:2017udc,Rastelli:2017ymc,Zhou:2017zaw,Zhou:2018ofp} for bootstrap-like methods of computing holographic correlators which does not require the detailed knowledge of the effective Lagrangian. See also \cite{Arutyunov:2017dti,Arutyunov:2018neq,Arutyunov:2018tvn} for an improved version of the original algorithm and \cite{Caron-Huot:2018kta} for a method based on the inversion formula \cite{Caron-Huot:2017vep} and crossing symmetry.} In particular an interesting $SO(10,2)$ symmetry \cite{Caron-Huot:2018kta}  was shown to exist in the general formula for one-half BPS four-point functions from tree-level IIB supergravity in $AdS_5\times S^5$ \cite{Rastelli:2016nze,Rastelli:2017udc}, unifiying the correlators of all Kaluza-Klein modes into a single ten-dimensional object. The simplicity in these results requires remarkable conspiracy of individual Witten diagrams, and some of its aspects still remain to be better understood.  Finally, one is led to study Witten diagrams from the reconstruction of AdS physics using CFT principles \cite{Heemskerk:2009pn}. Witten diagrams emerge as solutions to the crossing equation, at tree level \cite{Heemskerk:2009pn,Alday:2017gde,Li:2017lmh} as well as at  loop level \cite{Aharony:2016dwx}. 
 
 However, some basic properties of Witten diagrams still remain to be better understood, even at tree level. In particular, the following seemingly simple problem still appears to lack a satisfactory solution: how do we perform the conformal block decomposition of a tree-level exchange Witten diagram in the crossed channel? In the crossed channel an exchange Witten diagram is known to decompose into two towers of double-trace conformal blocks.\footnote{Here we assume that the external conformal dimensions are generic such that the spectra of the two towers of double-trace operators do not overlap. In the degenerate case, both conformal blocks and their derivative with respect to the conformal dimension will appear in the expansion. We will have more discussions on this point later in the paper.} However obtaining the crossed channel OPE coefficients turns out to be very non-trivial, and no method is currently available to extract efficiently {\it all} the coefficients. This should be contrasted with the decomposition in the direct channel, where it contains a single-trace conformal block and infinitely many double-trace blocks with bounded spins. The problem in the direct channel is much easier and can be solved using a variety of methods. For example one can obtain all the decomposition coefficients in closed forms using the split representation of propagators \cite{Costa:2014kfa}, or the geodesic Witten diagram techniques \cite{Hijano:2015zsa}.\footnote{In Section \ref{Secdirectchan} we will offer another approach to obtain these coefficients from studying the contact diagrams related to the exchange diagram.}

 A number of recent papers have appeared that revisit this problem, and the methods have both advantages and disadvantages. In  \cite{Liu:2018jhs,Cardona:2018dov} methods based on the Lorentizian inversion formula \cite{Caron-Huot:2017vep} (see also \cite{Simmons-Duffin:2017nub,Kravchuk:2018htv}) are introduced, which allow one to extract the crossed channel CFT data for operators with spins greater than the spin of the exchanged single-trace operator. However the inversion formula is not valid for lower spins due to the Regge behavior of the Witten diagrams, and one therefore cannot use this method to probe the rest of the operators. This difficulty is absent in Mellin space \cite{Mack:2009mi,Penedones:2010ue}. The OPE coefficients of the double-trace operators with the leading conformal twist can be obtained by taking the residue of the Mellin amplitude at the leading double-trace pole, and then projecting the residue into  continuous Hahn polynomials of different spins \cite{Costa:2014kfa,Gopakumar:2016cpb,Sleight:2018epi,Sleight:2018ryu,Gopakumar:2018xqi} -- analogues to projecting flat space amplitudes into partial waves using Gegenbaur polynomials. But in this approach one encounters a different difficulty in obtaining the OPE coefficients of double-trace operators with sub-leading twists. The residue of the Mellin amplitude at a sub-leading double-trace pole receives contributions from both the double-trace primary operators, as well as the conformal descendants of the double-trace operators whose twists are smaller. Therefore there is a mixing of contributions between the two. The degeneracy problem must be first solved in order to extract the OPE coefficients. One method of disentangling the contributions was suggested in \cite{Sleight:2018epi,Sleight:2018ryu}.   
In this method one first acts on the correlator with a quartic differential operator and then repeats the Mellin space analysis of performing the projections on the residue. The differential operator can be chosen such that its kernel contains the conformal blocks of the leading double-trace twist  \cite{Alday:2016njk}, the sub-leading double-trace operators therefore become leading in the new correlator. Unfortunately, the complexity of this algorithm quickly grows upon increasing the twists of the double-trace operators. Applying this method to low orders also reveals few general patterns.

In this paper, we will attack the problem from a different angle. The main result of our analysis is  a set of simple constraining linear relations satisfied by the crossed channel decomposition coefficients. These linear relations will lead us to a recursive algorithm for solving the coefficients. In our analysis we emphasize the pivotal role played by the contact Witten diagrams, and highlight the importance of an ``equation of motion'' operator which relates an exchange Witten diagram to a sum of contact Witten diagrams. More precisely, the equation of motion operator is given by the quadratic conformal Casimir operator in the exchange channel with a constant shift, and is closely related to the equation of motion for the bulk-to-bulk propagator. The direct consequence of this relation between these two types of diagrams is that the direct channel decomposition coefficients of an exchange Witten diagram are completely fixed in terms of the coefficients of the related contact Witten diagrams. This relation has important implications in the crossed channel too. Using properties of conformal blocks \cite{Dolan:2011dv}\footnote{See also \cite{Karateev:2017jgd}.}, we find that the equation of motion operator admits simple actions on conformal blocks. In one dimension, the action of this operator on a conformal block with dimension $\Delta$ produces three conformal blocks with new dimensions $\Delta-1$, $\Delta$ and $\Delta+1$. In higher dimensions, the three-term relation becomes a five-term one. The action of the  operator  on a conformal block with dimension $\Delta$ and spin $\ell$  contains the original conformal block, as well as four other conformal blocks with shifted quantum numbers $(\Delta\mp1,\ell\pm1)$, $(\Delta\pm1,\ell\mp1)$. Moreover, the coefficients of the three-term and five-term relations vanish for conformal blocks of unphysical double-trace operators\footnote{These are the operators with conformal dimensions $\Delta=\Delta_1+\Delta_2+\ell-1, \Delta_3+\Delta_4+\ell-1$ or negative spins $\ell=-1$ for $d>1$, and $\Delta=\Delta_1+\Delta_2-1,\Delta_3+\Delta_4-1$ for $d=1$.}. These zeros of the coefficients guarantee that  when we restrict $(\Delta,\ell)$ to be those of the double-trace operators, the spectrum is preserved after applying these relations. The simple action of the equation of motion operator on double-trace conformal blocks makes it possible to formulate a recursive algorithm for solving the crossed channel decomposition coefficients. Thanks to the equation of motion identity, these crossed channel coefficients satisfy simple linear equations with the decomposition coefficients of the contact Witten diagrams as inhomogeneous terms. We can solve the linear equations recursively, in terms of certain seed decomposition coefficients. In one dimension, the seed coefficients are just the OPE coefficients of the double-trace operators with the lowest conformal dimension. In higher dimensions, the seed coefficients are the OPE coefficients of the leading twist double-trace operators. Therefore the recursion relations give us a very efficient way to obtain OPE coefficients of sub-leading double-trace operators. 

The decomposition of exchange Witten diagrams in the crossed channel is also closely related to the $6j$ symbol (or the crossing kernel) of the conformal group. In the crossed channel, a conformal partial wave is decomposed into infinitely many double-trace conformal blocks. The various decomposition coefficients of the double-trace operators can be viewed as the residues of the $6j$ symbol  \cite{Liu:2018jhs}. Because a conformal partial wave can be identified with the difference of two exchange Witten diagrams with opposite quantizations \cite{Penedones:2007ns,Costa:2014kfa,Giombi:2018vtc}, our analysis of the Witten diagrams extends easily to conformal partial waves.

The paper is organized as follows. We start by defining the Witten diagrams in Section \ref{SecWD} and introducing the equation of motion operator in \ref{SecExtoCon}. In Section \ref{CPW} we review some basic facts of conformal partial waves. In Section \ref{SecCBdecompContact} we discuss the conformal block decomposition of contact Witten diagrams. We show that the decomposition of a generic contact diagram can always be recursively reduced to the simplest contact diagram with zero derivatives in the quartic vertex. In Section \ref{Secdirectchan} we show how the direct channel decomposition of exchange Witten diagrams can be fixed by the decomposition of the relevant contact Witten diagrams. In Section \ref{SecRecurcross}, we discuss the crossed channel decomposition of exchange Witten diagrams and conformal partial waves. We first outline the strategy in Section \ref{SecStrat}. We discuss the simpler problem in $\mathrm{CFT}_1$ in Section \ref{1d}, and then extend the story to $\mathrm{CFT}_d$ in Section \ref{higherd}. We end with a brief discussion in Section \ref{SecDiscuss}. Further technical details are relegated to the three appendices. In Appendix \ref{appcontact} we make further comments on the contact Witten diagrams. In Appendix \ref{appeqweight} we discuss the special case where the external conformal dimensions are degenerate. In Appendix \ref{appseed} we discuss how to compute the seed coefficients for $AdS_2$ exchange Witten diagrams.

\section{Witten Diagrams and Conformal Partial Waves}
\subsection{Witten Diagrams}\label{SecWD}
In this paper we study tree-level four-point Witten diagrams, {\it i.e.}, contact Witten diagrams and exchange Witten diagrams. We focus on scalar Witten diagrams where the external operators have zero spins and conformal dimensions $\Delta_i$, $i=1,2,3,4$. This restricts the operator in the internal line of the exchange Witten diagram to be in the rank-$\ell_E$ symmetric-traceless representation under the Lorentz group. We will denote the conformal dimension of the exchange operator by $\Delta_E$. After making a choice for the cubic and quartic vertices, the Witten diagrams are built from the scalar bulk-to-boundary propagators $G_{B\partial}^{\Delta_i}(z,\vec{x}_i)$, and the spin-$\ell_E$ bulk-to-bulk propagator $\Pi^{\Delta_E}_{\mu_1\ldots \mu_{\ell_E},\nu_1\ldots \nu_{\ell_E}}(z_1,z_2)$. The scalar bulk-to-boundary propagator is explicitly given by \begin{equation}
G_{B\partial}^{\Delta_i}(z,\vec{x}_i)=\left(\frac{z_0}{z_0^2+(\vec{z}-\vec{x}_i)^2}\right)^{\Delta_i}
\end{equation}
where the dimension $\Delta_i$ is associated to the scalar field mass in $AdS_{d+1}$ via $M_i^2=\Delta_i(\Delta_i-d)$. The spin-$\ell_E$ bulk-to-bulk propagators are defined to satisfy the equation of motion and have vanishing divergence
\begin{equation}
(\bigtriangledown_1^2-M_E^2)\,\Pi^{\Delta_E}_{\mu_1\ldots \mu_{\ell_E},\nu_1\ldots \nu_{\ell_E}}(z_1,z_2)=-g^{\mu_1\{\nu_1}\ldots g^{|\mu_{\ell_E}|\nu_{\ell_E}\}}\delta(z_1,z_2)+\ldots\;,\label{EOMofBtoBprop}
\end{equation}
\begin{equation}
\bigtriangledown_1^{\mu_1}\,\Pi^{\Delta_E}_{\mu_1\ldots \mu_{\ell_E},\nu_1\ldots \nu_{\ell_E}}(z_1,z_2)=0+\ldots\;,\label{divergenceless}
\end{equation}
up to local source terms denoted by $\ldots$. These terms introduce ambiguities to the exchange Witten diagrams\footnote{These terms will only change the exchange Witten diagram by contact Witten diagrams \cite{Costa:2014kfa}.}, but such ambiguities are not important for the propagating degrees of freedom. The squared mass of the bulk field is given by
\begin{equation}
M_E^2=\Delta_E(\Delta_E-d)-\ell_E\;.
\end{equation}
An explicit example of the bulk-to-bulk propagator is given by that of  a scalar field
\begin{equation}\small
\Pi^{\Delta_E}(z_1,z_2)=\frac{\Gamma(\Delta_E)}{2\pi^{\frac{d}{2}}\Gamma(\Delta_E-\frac{d}{2}+1)}u^{-\Delta}{}_2F_1\left(\Delta_E,\frac{2\Delta_E-d+1}{2},2\Delta_E-d+1,-\frac{4}{u}\right)
\end{equation}
where
\begin{equation}
u=\frac{(z_1-z_2)^2}{z_{10}z_{20}}\;.
\end{equation}

A generic contact Witten diagram is built in terms of the bulk-to-boundary propagators only
\begin{equation}\label{Wcontact}
W^{contact}=\int \frac{d^{d+1}z}{z_0^{d+1}}\prod_{i=1}^4 (\bigtriangledown^{\mu})^{j_i}G^{\Delta_i}_{B\partial}(z,x_i)\;.
\end{equation}
In this formula, $(\bigtriangledown^{\mu})^{j_i}$ is schematic for the product of covariant derivatives $\bigtriangledown^{\mu_1}\ldots \bigtriangledown^{\mu_{j_i}}$, and the indices are appropriately contracted on the RHS of (\ref{Wcontact}). Such a contact diagram arises from a quartic AdS contact vertex
\begin{equation}
(\bigtriangledown^{\mu})^{j_1}\phi_1(\bigtriangledown^{\mu})^{j_2}\phi_2(\bigtriangledown^{\mu})^{j_3}\phi_3(\bigtriangledown^{\mu})^{j_4}\phi_4
\end{equation}
which contains a total number of $j_1+j_2+j_3+j_4$ derivatives. When $j_1=j_2=j_3=j_4=0$, we have the simplest contact diagram which is denoted in the literature as a $D$-function
\begin{equation}\label{Dfunction}
D_{\Delta_1\Delta_2\Delta_3\Delta_4}=\int \frac{d^{d+1}z}{z_0^{d+1}}\prod_{i=1}^4 G^{\Delta_i}_{B\partial}(z,x_i)\;.
\end{equation} 

We also define a spin-$\ell_E$ exchange Witten diagram (in the s-channel) as
\begin{equation}\label{sWitten}
\begin{split}
W^{s,\, exchange}_{\Delta_E,\ell_E}={}&\int \frac{d^{d+1}z_1}{z_{10}^{d+1}}\frac{d^{d+1}z_2}{z_{20}^{d+1}}G^{\Delta_1}_{B\partial}(z_1,x_1)((\bigtriangledown^{\mu})^{\ell_E}G^{\Delta_2}_{B\partial}(z_1,x_2))\Pi^{\Delta_E}_{\mu_1\ldots \mu_{\ell_E},\nu_1\ldots \nu_{\ell_E}}(z_1,z_2)\\
{}&\times G^{\Delta_3}_{B\partial}(z_2,x_3) ((\bigtriangledown^{\nu})^{\ell_E}G^{\Delta_4}_{B\partial}(z_2,x_4))
\end{split}
\end{equation}
where we assumed the cubic couplings are 
\begin{equation}
\phi_1\bigtriangledown^{\mu_1}\ldots\bigtriangledown^{\mu_{\ell_E}}\phi_2 h_{\mu_1\ldots\mu_{\ell_E}}\;,\quad\text{and}\quad \phi_1\bigtriangledown^{\mu_1}\ldots\bigtriangledown^{\mu_{\ell_E}}\phi_2 h_{\mu_1\ldots\mu_{\ell_E}}\;.
\end{equation}
Other distributions of the derivatives in the cubic vertices can be obtained from the above choice by using integration by parts. Because of (\ref{EOMofBtoBprop}) and (\ref{divergenceless}), the different choices of the cubic vertices will only affect $W^{s,\, exchange}$ by a finite number of contact diagrams.

The main goal of this paper is to study the conformal block decomposition of these diagrams. From the Mellin representation of Witten diagrams \cite{Penedones:2010ue,Paulos:2011ie,Fitzpatrick:2011ia,Costa:2012cb}, it is clear that a contact Witten diagram (\ref{Wcontact}) decomposes only into double-trace operators. For example, in the s-channel the decomposition reads\footnote{In this paper we will abuse the terminology by calling the coefficients in front of the conformal blocks, such as $a^{12}_{n,J}$, $a^{34}_{n,J}$, the ``OPE coefficients''. }
\begin{equation}\label{Wcontactins}
W^{contact}(x_i)=\sum_{J=0}^{J_{\rm max}}\sum_{n=0}^\infty a^{12}_{n,J} g^{(s)}_{\Delta_1+\Delta_2+2n+J,J}(x_i)+\sum_{J=0}^{J_{\rm max}}\sum_{n=0}^\infty a^{34}_{n,J} g^{(s)}_{\Delta_3+\Delta_4+2n+J,J}(x_i)\;.
\end{equation}
Notice that the spin of the conformal blocks has a finite support $0\leq J\leq J_{\rm max}\leq j_1+j_2+j_3+j_4$. 

Decomposing an exchange Witten diagram into the direct channel, one finds a single-trace conformal block and infinitely many double-trace blocks
\begin{equation}\label{Wexchangeins}
W^{s,\,exchange}_{\Delta_E,\ell_E}=A\, g^{(s)}_{\Delta_E,\ell_E}(x_i) +\sum_{J=0}^{\ell_E}\sum_{n=0}^\infty A^{12}_{n,J} g^{(s)}_{\Delta_1+\Delta_2+2n+J,J}(x_i)+\sum_{J=0}^{\ell_E}\sum_{n=0}^\infty A^{34}_{n,J} g^{(s)}_{\Delta_3+\Delta_4+2n+J,J}(x_i)\;.
\end{equation}
Again, the support of spins is finite. In contrast, when we decompose an exchange Witten diagram into the crossed channel ({\it i.e.}, the t-channel and u-channel), we find only double-trace operators of which the spins are unbounded
\begin{equation}\label{Wexchangeint}
W^{s,\,exchange}_{\Delta_E,\ell_E}=\sum_{J=0}^{\infty}\sum_{n=0}^\infty B^{14}_{n,J} g^{(t)}_{\Delta_1+\Delta_4+2n+J,J}(x_i)+\sum_{J=0}^{\infty}\sum_{n=0}^\infty B^{23}_{n,J} g^{(t)}_{\Delta_2+\Delta_3+2n+J,J}(x_i)\;.
\end{equation}
The above discussion is for {\it generic} external conformal dimensions. When $\Delta_1+\Delta_2-\Delta_3-\Delta_4\in 2\mathbb{Z}$, we will also encounter derivative conformal blocks $\partial_\Delta g^{(s)}_\Delta(x_i)$ in the s-channel.\footnote{\label{FNderivativeblock}We can understand this fact from just large $N$ counting. For concreteness we use the counting of 4d $\mathcal{N}=4$ SYM, then the tree level Witten diagrams are all of order $\mathcal{O}(1/N^2)$. If $\Delta_1+\Delta_2- \Delta_3-\Delta_4\neq 2\mathbb{Z}$, there is no order $\mathcal{O}(1)$ overlap between the double-trace spectra of operators $:O_1\square^{n_{12}}\partial^{\ell_{12}}O_2:$ and operators $:O_3\square^{n_{34}}\partial^{\ell_{34}}O_4:$. The double trace operators $:O_1\square^{n_{12}}\partial^{\ell_{12}}O_2:$ therefore can only appear in the $O_3\times O_4$ OPE with a suppression power of $1/N^2$ (the same for $:O_3\square^{n_{34}}\partial^{\ell_{34}}O_4:$ to appear in the $O_1\times O_2$ OPE). The conformal dimensions of the double-trace operators are corrected at order $\mathcal{O}(1/N^2)$. But their effect is invisible in the tree diagrams because the correction is of order $\mathcal{O}(1/N^4)$ due to the $\mathcal{O}(1/N^2)$ suppression in the OPE coefficients. On the other hand, when $\Delta_1+\Delta_2- \Delta_3-\Delta_4$ is an even integer (which we can further assume to be non negative), the double-trace operators with twists $\tau\geq\Delta_1+\Delta_2$ appear in both OPEs with $\mathcal{O}(1)$ coefficients. The correction due to the anomalous dimensions of these operators are therefore now visible. We can view the appearance of derivative double-trace blocks $\partial g^{(s)}_{\Delta_1+\Delta_2+2n+J,J}$ as expanding the anomalous dimension in $g^{(s)}_{\Delta_1+\Delta_2+2n+J+\frac{1}{N^2}\gamma_{n,J},J}$ to $\mathcal{O}(1/N^2)$.
 } Similarly,  when $\Delta_1+\Delta_4-\Delta_2-\Delta_3$ or $\Delta_1+\Delta_3-\Delta_2-\Delta_4$ is an even integer, there will be derivative conformal blocks in the t or u-channel. 

For $d=1$, the decomposition of Witten diagrams has the same qualitative features, except that there is no spin. In the s-channel decomposition, we have
\begin{equation}\label{Wcontactins1d}
W^{contact}(x_i)=\sum_{n=0}^\infty a^{12}_{n} g^{(s)}_{\Delta_1+\Delta_2+2n}(x_i)+\sum_{n=0}^\infty a^{34}_{n} g^{(s)}_{\Delta_3+\Delta_4+2n}(x_i)\;,
\end{equation}
\begin{equation}\label{Wexchangeins1d}
W^{s,exchange}(x_i)=Ag^{(s)}_\Delta(x_i)+\sum_{n=0}^\infty A^{12}_{n} g^{(s)}_{\Delta_1+\Delta_2+2n}(x_i)+\sum_{n=0}^\infty A^{34}_{n} g^{(s)}_{\Delta_3+\Delta_4+2n}(x_i)\;,
\end{equation}
and only double-trace operators with one ``parity'' will show up, {\it i.e.}, $\Delta_1+\Delta_2+n$ and $\Delta_3+\Delta_4+n$ with $n$ even. In the t-channel decomposition of the exchange Witten diagram,
\begin{equation}\label{Wexchangeint1d}
W^{s,\,exchange}_{\Delta_E}=\sum_{n=0}^\infty B^{14}_{n} g^{(t)}_{\Delta_1+\Delta_4+n}(x_i)+\sum_{n=0}^\infty B^{23}_{n} g^{(t)}_{\Delta_2+\Delta_3+n}(x_i)\;,
\end{equation}
double-trace operators of both parities will appear, {\it i.e.}, $n\in \mathbb{Z}$.

\subsection{Relating Exchange Diagrams to Contact Diagrams}\label{SecExtoCon}
The exchange Witten diagrams are related  to the contact Witten diagrams by the a second order differential operator, as a result of the fact that the bulk-to-bulk propagators are Green's functions in AdS. To see this explicitly, let us first focus on the $z_1$ integral inside
 the s-channel exchange Witten diagram (\ref{sWitten})\begin{equation}
I^{s,\,exchange}_{\nu_1\ldots\nu_{\ell_E}}(x_1,x_2;z_2)=\int \frac{d^{d+1}z_1}{z_{10}^{d+1}}G^{\Delta_1}_{B\partial}(z_1,x_1)((\bigtriangledown^{\mu})^{\ell_E}G^{\Delta_2}_{B\partial}(z_1,x_2))\Pi^{\Delta_E}_{\mu_1\ldots \mu_{\ell_E},\nu_1\ldots \nu_{\ell_E}}(z_1,z_2)\;.
\end{equation}
This integral is manifestly invariant under $SO(d,2)$, and therefore satisfies the identity 
\begin{equation}
(\mathbf{L}_1+\mathbf{L}_2+\mathfrak{L}_{z_2})_{AB}\,I^{s,\,exchange}_{\nu_1\ldots\nu_{\ell_E}}(x_1,x_2;z_2)=0\;.
\end{equation}
Here $\mathbf{L}_1$ and $\mathbf{L}_2$ are the conformal generators which act on $x_1$ and $x_2$, and $\mathfrak{L}_{z_2}$ is the $AdS_{d+1}$ isometry generator which acts on a spin-$\ell_E$ field at $z_2$. Using this identity twice, we obtain the following action of the conformal Casimir operator with respect to $x_1$ and $x_2$
\begin{equation}
-\frac{1}{2}(\mathbf{L}_1+\mathbf{L}_2)^2\,I^{s,\,exchange}_{\nu_1\ldots\nu_{\ell_E}}=-\frac{1}{2}\mathfrak{L}_{z_2}^2\,I^{s,\,exchange}_{\nu_1\ldots\nu_{\ell_E}}=(\bigtriangledown^2_{z_2}+\ell_E(\ell_E+d-1))I^{s,\,exchange}_{\nu_1\ldots\nu_{\ell_E}}\;.
\end{equation}
In the second equality, we have used that the conformal Casimir is equal to the Laplacian up to a constant shift $\ell_E(\ell_E+d-1)$ \cite{Pilch:1984xx}. We now apply the equation of motion (\ref{EOMofBtoBprop}) to get rid of $\bigtriangledown^2_{z_2}$ that acts on the bulk-to-bulk propagator, and perform the $z_2$ integral. We find the expression gives the following relation between an exchange diagram and a sum of contact Witten diagrams
\begin{equation}\label{EOMWexWcon}
\left[\frac{1}{2}(\mathbf{L}_1+\mathbf{L}_2)^2+C^{(2)}_{\Delta_E,\ell_E}\right]W^{s,\, exchange}_{\Delta_E,\ell_E}=\sum_I c_IW^{contact}_I\;.
\end{equation}
Here $\sum_I c_IW^{contact}_I$ is a collection of contact Witten diagrams, obtained by replacing the bulk-to-bulk propagator with the RHS of (\ref{EOMofBtoBprop}).\footnote{One might worry about the contact term ambiguities in defining an exchange Witten diagram, introduced by the ``\ldots'' in (\ref{EOMofBtoBprop}) and (\ref{divergenceless}). What is the action of the operator $\frac{1}{2}(\mathbf{L}_1+\mathbf{L}_2)^2+C^{(2)}_{\Delta_E,\ell_E}$ on a contact Witten diagram? It turns out that the action of the equation of motion operator on a contact Witten can again be expressed as the linear combination of finitely many contact Witten diagrams, as we will show at the end of Appendix \ref{appcontact}. Therefore the relation (\ref{EOMWexWcon}) holds for exchange Witten diagrams independent of the choice of contact terms. \label{footnoteEOMonContact}} $C^{(2)}_{\Delta_E,\ell_E}$ is the eigenvalue of the quadratic conformal Casimir for an operator with dimension $\Delta_E$ and spin $\ell_E$
\begin{equation}
C^{(2)}_{\Delta_E,\ell_E}=\Delta_E(\Delta_E-d)+\ell_E(\ell_E+d-2)\;.
\end{equation}

Let us work out how the operator on the LHS acts as a differential operator in terms of the cross ratios. It is standard to extract a kinematic factor from the four-point function
\begin{equation}\label{GUV}
\langle O_1(x_1)\ldots O_4(x_4) \rangle\equiv G(x_i)=\frac{1}{(x_{12}^2)^{\frac{\Delta_1+\Delta_2}{2}}(x_{34}^2)^{\frac{\Delta_3+\Delta_4}{2}}}\left(\frac{x_{14}^2}{x_{24}^2}\right)^{a}\left(\frac{x_{14}^2}{x_{13}^2}\right)^{b}\mathcal{G}(U,V)
\end{equation}
such that it becomes a function of the conformal cross ratios  $U$ and $V$
\begin{equation}
U=\frac{x_{12}^2x_{34}^2}{x_{13}^2x_{24}^2}\;,\quad V=\frac{x_{14}^2x_{23}^2}{x_{13}^2x_{24}^2}\;.
\end{equation}
We have also defined
\begin{equation}\small
a=\frac{\Delta_2-\Delta_1}{2}\;,\quad b=\frac{\Delta_3-\Delta_4}{2}\;.
\end{equation}
The action of $\frac{1}{2}(\mathbf{L}_1+\mathbf{L}_2)^2+C^{(2)}_{\Delta_E,\ell_E}$ on a function of $x_i$ defines an operator acting on $\mathcal{G}(U,V)$. Let us denote this operator as $\mathbf{EOM}^{(s)}$, then (\ref{EOMWexWcon}) becomes
\begin{equation}\label{EOMWexWconcrossratio}
\mathbf{EOM}^{(s)}[\mathcal{W}^{s,\, exchange}_{\Delta_E,\ell_E}(U,V)]=\sum_I c_I\mathcal{W}^{contact}_I(U,V)\;.
\end{equation}
 To give an explicit expression for this operator, it is convenient to make a change of variables for the conformal cross ratios
\begin{equation}
U=z\bar{z}\;,\quad V=(1-z)(1-\bar{z})\;,
\end{equation}
and define a second order differential operator
\begin{equation}
\mathbf{D}_{z}(a,b)=(1-z)z^2\frac{d^2}{dz^2}-(1+a+b)z^2\frac{d}{dz}-abz\;.
\end{equation}
The s-channel equation of motion operator can be written as
\begin{equation}\label{EOMszzb}
\mathbf{EOM}^{(s)}[\mathcal{G}(z,\bar{z})]=-2\mathbf{\Delta}_\epsilon(a,b)[\mathcal{G}(z,\bar{z})]+C^{(2)}_{\Delta_E,\ell_E}\, \mathcal{G}(z,\bar{z})
\end{equation}
where
\begin{equation}
\mathbf{\Delta}_\epsilon(a,b)=\mathbf{D}_z(a,b)+\mathbf{D}_{\bar{z}}(a,b)+2\epsilon \frac{z\bar{z}}{z-\bar{z}}\bigg((1-z)\frac{d}{dz}-(1-\bar{z})\frac{d}{d\bar{z}}\bigg)\;,
\end{equation}
and we have defined
\begin{equation}
\epsilon=\frac{d}{2}-1\;.
\end{equation}
Note that the s-channel conformal blocks are eigenfunctions of this differential operator \cite{Dolan:2003hv} \footnote{In this paper we slightly abuse the notation to let $g^{(s)}_{\Delta,\ell}(z,\bar{z})$ also denote the conformal block as a function of the cross ratios where a kinematic factor of $x_{ij}^2$ is extracted from $g^{(s)}_{\Delta,\ell}(x_i)$ according to (\ref{GUV}). The meaning should be clear from their different arguments and the context.}
\begin{equation}
\mathbf{EOM}^{(s)}[g^{(s)}_{\Delta,\ell}(z,\bar{z})]=(C^{(2)}_{\Delta,\ell}-C^{(2)}_{\Delta_E,\ell_E})g^{(s)}_{\Delta,\ell}(z,\bar{z})\;.
\end{equation}
In particular, the operator annihilates the single-trace block $g^{(s)}_{\Delta_E,\ell_E}(z,\bar{z})$.

For $d=1$, the bulk space is $AdS_2$ and only scalar fields propagate in the internal line of the exchange Witten diagrams. Acting with the Casimir operator on the scalar exchange Witten diagram, we get the $D$-function (\ref{Dfunction})
\begin{equation}
\left[\frac{1}{2}(\mathbf{L}_1+\mathbf{L}_2)^2+\Delta_E(\Delta_E-1)\right]W^{s,\,exchange}_{\Delta_E}=D_{\Delta_1\Delta_2\Delta_3\Delta_4}\;.
\end{equation}
Moreover, because there is only one cross ratio in one dimension
\begin{equation}
z=\frac{x_{12}x_{34}}{x_{13}x_{24}}\;,
\end{equation}
the operator $\mathbf{EOM}^{(s)}$ is a differential operator of $z$, and is given by
\begin{equation}
\mathbf{EOM}^{(s)}=-\mathbf{D}_{z}(2a,2b)+\Delta_E(\Delta_E-1)\;.
\end{equation}

\subsection{Conformal Partial Waves}\label{CPW}
It is sometimes useful to think of the conformal block decomposition of conformal correlators as arising from a more primitive formula, in terms of the so-called conformal partial waves $\Psi_{\Delta,J}^{(s)}(x_i)$.  We review in this section some basic properties of conformal partial waves for reader's convenience.

The conformal partial wave takes the form as the sum of the conformal blocks and its shadow block with $\Delta\to\widetilde{\Delta}=d-\Delta$
\begin{equation}\label{CPWdef}
\Psi_{\Delta,J}^{(s)}(x_i)=K^{\Delta_3,\Delta_4}_{\widetilde{\Delta},J}g^{(s)}_{\Delta,J}(x_i)+K^{\Delta_1,\Delta_2}_{\Delta,J}g^{(s)}_{\widetilde{\Delta},J}(x_i)
\end{equation}
where the coefficients are given by
\begin{equation}
K^{\Delta_1,\Delta_2}_{\Delta,J}=\left(-\frac{1}{2}\right)^J\frac{\pi^{\frac{d}{2}}\Gamma(\Delta-\frac{d}{2})\Gamma(\Delta+J-1)\Gamma(\frac{\widetilde{\Delta}+\Delta_1-\Delta_2+J}{2})\Gamma(\frac{\widetilde{\Delta}+\Delta_2-\Delta_1+J}{2})}{\Gamma(\Delta-1)\Gamma(d-\Delta+J)\Gamma(\frac{\Delta+\Delta_1-\Delta_2+J}{2})\Gamma(\frac{\Delta+\Delta_2-\Delta_1+J}{2})}\;.
\end{equation}
Conformal partial waves with all integer spins $J$ and unphysical complex dimensions $\Delta=\frac{d}{2}+i \nu$, $\nu\geq0$, form a complete set of functions \cite{Dobrev:1977qv}, and are usually referred to as the principal series representation. In terms of conformal partial waves, a four-point correlation function can be written as a contour integral \cite{Dobrev:1975ru}\footnote{More precisely, for $d=1$ the principal series is also supplemented by a discrete series with $\Delta=2n$, $n=1,2,\ldots$. Here we assume that the spectral function $\rho(\Delta)$ does not contain physical poles at $\Delta=2n$. Then after closing the contour and taking the residues, poles from the conformal partial wave will precisely cancel the contribution from the discrete series.}
\begin{equation}
\begin{split}
G(x_i)={}&\sum_{J=0}^{\infty}\int_{\frac{d}{2}}^{\frac{d}{2}+i\infty} \frac{d\Delta}{2\pi i}\,\rho(\Delta,J)\,\Psi_{\Delta,J}^{(s)}(x_i)\\
={}&\sum_{J=0}^{\infty}\int_{\frac{d}{2}}^{\frac{d}{2}+i\infty} \frac{d\Delta}{2\pi i}\,\rho(\Delta,J)\,(K^{\Delta_3,\Delta_4}_{\widetilde{\Delta},J}g^{(s)}_{\Delta,J}(x_i)+K^{\Delta_1,\Delta_2}_{\Delta,J}g^{(s)}_{\widetilde{\Delta},J}(x_i))
\end{split}
\end{equation}
Using the symmetry that 
\begin{equation}
\rho(\Delta,J)\,K^{\Delta_3,\Delta_4}_{\widetilde{\Delta},J}=\rho(\widetilde{\Delta},J)K^{\Delta_1,\Delta_2}_{\widetilde{\Delta},J}\;,
\end{equation}
we can eliminate the shadow blocks from the above integral and rewrite it as
\begin{equation}
G(x_i)=\sum_{J=0}^{\infty}\int_{\frac{d}{2}-i\infty}^{\frac{d}{2}+i\infty} \frac{d\Delta}{2\pi i}\,\rho(\Delta,J)\, K^{\Delta_3,\Delta_4}_{\widetilde{\Delta},J}g^{(s)}_{\Delta,J}(x_i)\;.
\end{equation}
After closing the contour to the right and picking up the poles, we get the usual conformal block decomposition.

The combination (\ref{CPWdef}) of the conformal block with its shadow is special because it makes the conformal partial wave a single-valued function in Euclidean space ({\it i.e.}, when $\bar{z}=z^*$). By contrast, each individual conformal block is not\footnote{For example this can be explicitly seen from the 2d conformal block
\begin{equation}
g^{(s)}_{\Delta,J}(z,\bar{z})=\frac{k_{\Delta-J}(z)k_{\Delta+J}(\bar{z})+k_{\Delta+J}(z)k_{\Delta-J}(\bar{z})}{1+\delta_{J,0}}\;,\quad k_{\beta}(z)=z^{\beta/2}{}_2F_1(\beta/2+a,\beta/2+b,\beta;z)\;.
\end{equation}
Around $z=1$ and $\bar{z}=1$, we can use the property of ${}_2F_1$ to write the conformal block as
\begin{equation}
g^{(s)}_{\Delta,J}(z,\bar{z})=f_{1}(z,\bar{z})(z-1)^{-a-b}(\bar{z}-1)^{-a-b}+f_{2}(z,\bar{z})+f_{3}(z,\bar{z})(z-1)^{-a-b}+f_4(z,\bar{z})(\bar{z}-1)^{-a-b}
\end{equation}
where $f_i$ are regular at $z=\bar{z}=1$. In the Euclidean regime $\bar{z}=z^*$, the first two terms are single-valued while the latter two terms fail to be.

}. We can most easily see this single-valuedness of the conformal partial wave from its integral representation \cite{Ferrara:1972xe,Ferrara:1973vz,Ferrara:1972uq,Ferrara:1972ay,SimmonsDuffin:2012uy}
\begin{equation}\label{CPWintegralrep}
\Psi_{\Delta,J}^{(s)}(x_i)=\int d^dx^5\langle O_1(x_1)O_2(x_2)O_5^{\mu_1\ldots\mu_J}(x_5)\rangle \langle \widetilde{O}_{5,\mu_1\ldots\mu_J}(x_5)O_3(x_3)O_4(x_4)\rangle\;,
\end{equation}
where $\widetilde{O}_{5,\mu_1\ldots\mu_J}$ is the shadow operator of $O_5$. This integral is manifestly single-valued in Euclidean space. 

The above integral representation can also be lifted into the AdS space \cite{Costa:2014kfa}
\begin{equation}\label{AdSlift}
\begin{split}
\Psi_{\Delta,J}^{(s)}(x_i)\propto{}& \int d^d x_5 \int \frac{d^{d+1}z_1}{z_{10}^{d+1}} G^{\Delta_1}_{B\partial}(z_1,x_1)(\bigtriangledown^\mu)^JG^{\Delta_2}_{B\partial}(z_1,x_2)\Pi^{\Delta}_{\mu_1\ldots\mu_J}{}^{\rho_1\ldots\rho_J}(z_1,x_5)\\
{}& \times \int \frac{d^{d+1}z_2}{z_{20}^{d+1}} \Pi^{d-\Delta}_{\nu_1\ldots\nu_J,\rho_1\ldots\rho_J}(z_2,x_5) G^{\Delta_3}_{B\partial}(z_2,x_3)(\bigtriangledown^\nu)^J  G^{\Delta_4}_{B\partial}(z_2,x_4)
\end{split}
\end{equation}
where $\Pi^{\Delta}_{\mu_1\ldots\mu_J}{}^{\rho_1\ldots\rho_J}(z,x_5)$ is the spin-$J$ bulk-to-boundary propagator. It is clear that the first AdS integral over $z_1$ gives the three-point function $\langle O_1(x_1)O_2(x_2)O_5^{\mu_1\ldots\mu_J}(x_5)\rangle$ and the second AdS integral over $z_2$ gives $\langle \widetilde{O}_{5,\mu_1\ldots\mu_J}(x_5)O_3(x_3)O_4(x_4)\rangle$. Further performing the $x_5$ integral therefore reproduces the conformal partial wave $\Psi_{\Delta,J}^{(s)}(x_i)$ in the integral representation (\ref{CPWintegralrep}).

The above lift of the conformal block into AdS is closely related to the split representation of AdS propagators \cite{Costa:2014kfa}. It is convenient to first define the AdS harmonic function
\begin{equation}\label{Omegasplit}
\Omega^{\Delta,J}_{\mu_1\ldots\mu_J,\nu_1\ldots\nu_J}(z_1,z_2)=-\frac{(\Delta-\frac{d}{2})^2}{\pi J!(\frac{d}{2}-1)_J}\int d^dx_5 \Pi^{\Delta}_{\mu_1\ldots\mu_J}{}^{\rho_1\ldots\rho_J}(z_1,x_5)\Pi^{d-\Delta}_{\nu_1\ldots\nu_J,\rho_1\ldots\rho_J}(z_2,x_5)
\end{equation}
which splits a function of two bulk points into a product of two bulk-to-boundary propagators with a common  integrated boundary point. 
The bulk-to-bulk propagator can then be expanded in terms of these AdS harmonic functions
\begin{equation}\label{PiinOmega}
\Pi^{\Delta,J}_{\mu_1\ldots\mu_J,\nu_1\ldots\nu_J}(z_1,z_2)=\sum_{l=0}^J\int d\Delta' \,a_\ell(\Delta') (\bigtriangledown_1^\mu)^{J-\ell}(\bigtriangledown_2^\nu)^{J-\ell}\Omega^{\Delta',\ell}_{\mu_1\ldots\mu_l,\nu_1\ldots\nu_l}(z_1,z_2)\;
\end{equation}
where explicit coefficients $a_\ell(\Delta')$ can be found in \cite{Costa:2014kfa}. The above split representation of the AdS harmonic function (\ref{Omegasplit}) gives the split representation of the bulk-to-bulk propagator.

Relatedly, a conformal partial wave can also be represented in AdS as the difference of two exchange Witten diagrams with opposite quantizations
\begin{equation}\label{Psiasdifference}
\Phi^{(s)}_{\Delta,J}(x_i)\propto W^{s,exchange}_{\Delta,J}(x_i)-W^{s,exchange}_{d-\Delta,J}(x_i)\;.
\end{equation}
This fact was pointed out, {\it e.g.}, in \cite{Penedones:2007ns,Costa:2014kfa,Giombi:2018vtc}, and can be understood in two steps as follows. Firstly, all the double-trace conformal blocks in $W^{s,exchange}_{\Delta,J}$ and $W^{s,exchange}_{d-\Delta,J}$ cancel out in (\ref{Psiasdifference}). We can see this by noting that both $W^{s,exchange}_{\Delta,J}$ and $W^{s,exchange}_{d-\Delta,J}$ satisfy the same equation of motion identity
\begin{equation}
\mathbf{EOM}^{(s)}[\mathcal{W}^{s,\, exchange}_{\Delta_E,\ell_E}(U,V)]=\mathbf{EOM}^{(s)}[\mathcal{W}^{s,\, exchange}_{d-\Delta_E,\ell_E}(U,V)],
\end{equation}
because the contact diagrams are determined by vertices which are independent of the quantization. Moreover, the double-trace conformal blocks are diagonal under the equation of motion operator, with non vanishing and quantization-independent eigenvalues. This guarantees the cancellation of double-trace blocks in the difference and implies that (\ref{Psiasdifference}) is just the linear combination of the single-trace conformal block and its shadow. The second step is therefore to show that the linear combination is proportional to (\ref{CPWdef}). However, this is guaranteed by single-valuedness. Because the exchange Witten diagrams are single-valued in the Euclidean regime, there is only one way, {\it i.e.}, in the fashion of (\ref{CPWdef}), to combine the non-single-valued conformal block and its shadow in order to achieve single-valuedness .

As a final comment, let us mention that a single conformal block in $d\geq2$ {\it cannot} be decomposed into the crossed channel in terms of conformal blocks. This was noticed in, {\it e.g.}, \cite{ElShowk:2011ag,Liu:2018jhs}, and follows from the non single-valuedness of conformal block. In the Euclidean regime, an s-channel conformal block is not single-valued around $z=\bar{z}=1$ while the t-channel conformal blocks are. It is not possible to sum over infinitely many single-valued functions to obtain a non-single-valued function. On the other hand, a conformal partial wave can always be decomposed into the crossed channel in terms of double-trace conformal blocks. This follows from (\ref{Psiasdifference}) where the conformal partial wave is written as the difference of two s-channel exchange Witten diagrams, and each exchange Witten diagram admits decomposition into double-trace conformal blocks in the crossed channel. In 1d, the above comments do not apply. There is only one cross ratio and one can explicitly show that the conformal block can be decomposed into the crossed channel as infinitely many double-trace conformal blocks.

\section{Recursion Relations in Contact Witten Diagrams}
\subsection{Conformal Block Decomposition of Contact Witten Diagrams}\label{SecCBdecompContact}
In this subsection we focus on the conformal block decomposition of contact Witten diagrams (\ref{Wcontact}). We organize the contact Witten diagrams in terms of the total number of covariant derivatives in the quartic vertex. We start with the simplest contact Witten diagram where there is no derivative, {\it i.e.}, $D_{\Delta_1\Delta_2\Delta_3\Delta_4}$ defined in (\ref{Dfunction}).  The conformal block decomposition of higher-derivative contact Witten diagrams, as we will see, can be recursively related to the decomposition of the zero-derivative contact Witten diagrams. In this section, we will assume the external conformal dimensions are generic such that we will not encounter derivative conformal blocks. The conformal block decomposition for the special cases satisfying $\Delta_1+\Delta_2=\Delta_3+\Delta_4+2m$, {\it etc}, can be obtained from the generic case by taking the limit. We give more details of taking the limit in Appendix \ref{appcontact}.

\subsubsection*{The decomposition the zero-derivative contact diagram}
To obtain the conformal block decomposition of $D_{\Delta_1\Delta_2\Delta_3\Delta_4}$, we use a special case of the split representation, namely the split representation of the delta-function \cite{Penedones:2010ue,Costa:2014kfa,Bekaert:2015tva}
\begin{equation}
\delta(z_1,z_2)=\int d^dx_5\int_{-i\infty}^{i\infty} \frac{dc}{2\pi i} \,\rho_\delta(c)\,G_{B\partial}^{\frac{d}{2}+c}(z_1,x_5)\,G_{B\partial}^{\frac{d}{2}-c}(z_2,x_5)\;
\end{equation}
where 
\begin{equation}
\rho_\delta(c)=\frac{\Gamma(\frac{d}{2}+c)\Gamma(\frac{d}{2}-c)}{2\pi^d\Gamma(-c)\Gamma(c)}\;.
\end{equation}
Inserting this identity into (\ref{Dfunction}), we have 
\begin{equation}
D_{\Delta_1\Delta_2\Delta_3\Delta_4}=\int d^dx_5\int_{-i\infty}^{i\infty} \frac{dc}{2\pi i}\,\rho_\delta(c)\, D_{\Delta_1\,\Delta_2\, \frac{d}{2}+c}(x_1,x_2,x_5)D_{ \frac{d}{2}-c\, \Delta_3\,\Delta_4}(x_5,x_3,x_4)
\end{equation}
where $D_{\Delta_1\Delta_2\Delta_3}(x_1,x_2,x_3)$ is a three-point function
\begin{equation}\label{D3function}
\begin{split}
D_{\Delta_1\Delta_2\Delta_3}={}&\int \frac{d^{d+1}z}{z_0^{d+1}}\prod_{i=1}^3G^{\Delta_i}_{B\partial}(z,x_i)=\frac{a_{\Delta_1\Delta_2\Delta_3}}{x_{12}^{\Delta_1+\Delta_2-\Delta_3}x_{13}^{\Delta_1+\Delta_3-\Delta_2}x_{23}^{\Delta_2+\Delta_3-\Delta_1}}\;,\\
a_{\Delta_1\Delta_2\Delta_3}={}& \frac{\pi^{\frac{d}{2}}\Gamma(\frac{\Delta_1+\Delta_2-\Delta_3}{2})\Gamma(\frac{\Delta_1+\Delta_3-\Delta_2}{2})\Gamma(\frac{\Delta_2+\Delta_3-\Delta_1}{2})}{2\Gamma(\Delta_1)\Gamma(\Delta_2)\Gamma(\Delta_3)}\Gamma(\frac{\Delta_1+\Delta_2+\Delta_3-d}{2})\;.
\end{split}
\end{equation}
We can integrate out $x_5$ using (\ref{CPWintegralrep}), and obtain the conformal partial wave decomposition of the $D$-function
\begin{equation}\label{DfunctioninCPW}
D_{\Delta_1\Delta_2\Delta_3\Delta_4}=\int_{-i\infty}^{i\infty} \frac{dc}{2\pi i}\,\rho_D(c)\,\Psi^{(s)}_{\frac{d}{2}+c,0}(x_i)\;.
\end{equation}
The spectral density is
\begin{equation}
\rho_D(c)=\rho_\delta(c)\, a_{\Delta_1\,\Delta_2\,\frac{d}{2}+c}\,a_{\Delta_3\,\Delta_4\,\frac{d}{2}-c}\;.
\end{equation}
We can use the shadow symmetry in (\ref{DfunctioninCPW})to write it as a spectral representation with respect to the conformal blocks 
\begin{equation}
D_{\Delta_1\Delta_2\Delta_3\Delta_4}=\int_{-i\infty}^{i\infty} \frac{dc}{2\pi i}\,2\,\rho_D(c)\, K^{\Delta_3,\Delta_4}_{\frac{d}{2}-c,0}g^{(s)}_{\frac{d}{2}+c,0}(x_i)\;.
\end{equation}
By closing the contour to the right and taking the residues, we arrive at the conformal block decomposition of $D_{\Delta_1\Delta_2\Delta_3\Delta_4}$
\begin{equation}\label{Dfunctioningxi}
D_{\Delta_1\Delta_2\Delta_3\Delta_4}=\sum_{n=0}^\infty a^{12}_{n,0} g^{(s)}_{\Delta_1+\Delta_2+2n,0}(x_i)+\sum_{n=0}^\infty a^{34}_{n,0} g^{(s)}_{\Delta_3+\Delta_4+2n,0}(x_i)\;
\end{equation}
where 
\begin{equation}\label{Dfunctiondecomcoe}\small
\begin{split}
a^{12}_{n,0}={}&\frac{\pi ^{d/2} (-1)^{-n} \Gamma (n+\Delta_1) \Gamma (n+\Delta_2) \Gamma \left(-\frac{d}{2}+n+\Delta_1+\Delta_2\right) \Gamma \left(\frac{-d+2 n+\Delta_1+\Delta_2+\Delta_3+\Delta_4}{2}\right)}{2 n! \Gamma (\Delta_1) \Gamma (\Delta_2) \Gamma (\Delta_3) \Gamma (\Delta_4) \Gamma (2 n+\Delta_1+\Delta_2)}\\
{}&\times\frac{\Gamma \left(\frac{2 n+\Delta_1+\Delta_2+\Delta_3-\Delta_4}{2}\right) \Gamma \left(\frac{2 n+\Delta_1+\Delta_2-\Delta_3+\Delta_4}{2}\right) \Gamma \left(\frac{-2 n-\Delta_1-\Delta_2+\Delta_3+\Delta_4}{2}\right)}{ \Gamma \left(-\frac{d}{2}+2 n+\Delta_1+\Delta_2\right)}\;, 
\end{split}
\end{equation}
and $a^{34}_{n,0}$ can be obtained from $a^{12}_{n,0}$ by replacing $\Delta_1$, $\Delta_2$ with $\Delta_3$, $\Delta_4$.

\subsubsection*{The decomposition of higher-derivative contact diagrams}
Now let us consider a general contact Witten diagram (\ref{Wcontact}) with derivatives in the quartic vertex.  We first notice that a generic contact Witten diagram $W^{contact}$ can always be written as a linear combination of finitely many building blocks \cite{Penedones:2010ue}
\begin{equation}\label{Dnijdef}
D^{\{n_{ij}\}}_{\Delta_1\Delta_2\Delta_3\Delta_4}(x_i)\equiv\prod_{i<j}(x_{ij}^2)^{n_{ij}}D_{\Delta^{n_{ij}}_1\Delta^{n_{ij}}_2\Delta^{n_{ij}}_3\Delta^{n_{ij}}_4}(x_i)
\end{equation}
with {\it position-independent} coefficients. Here $\Delta^{n_{ij}}_k$ are the shifted dimensions
\begin{equation}
\Delta^{n_{ij}}_k=\Delta_k+\sum_{l\neq k}n_{kl}\;,
\end{equation}
and $n_{ij}=n_{ji}$ are {\it non negative} integers.  Note that under a conformal transformation, the function $D^{\{n_{ij}\}}_{\Delta_1\Delta_2\Delta_3\Delta_4}(x_i)$ transforms with conformal dimensions $\Delta^{n_{ij}}_k$. The conformal block decomposition of $W^{contact}$ is thus the {\it sum} of the decompositions of the building blocks $D^{\{n_{ij}\}}_{\Delta_1\Delta_2\Delta_3\Delta_4}$.

The second step is then to obtain the decomposition of $D^{\{n_{ij}\}}_{\Delta_1\Delta_2\Delta_3\Delta_4}$. Let us rewrite it in a different form
\begin{equation}\label{buildingblockandDshifted}
D^{\{n_{ij}\}}_{\Delta_1\Delta_2\Delta_3\Delta_4}=\frac{P(x_i,\Delta_1,\Delta_2,\Delta_3,\Delta_4)}{P(x_i,\Delta_1^{n_{ij}},\Delta_2^{n_{ij}},\Delta_3^{n_{ij}},\Delta_4^{n_{ij}})}(U^{-\frac{1}{2}}V)^{n_{23}}(U^{-\frac{1}{2}})^{n_{13}+n_{14}+n_{24}}D_{\Delta^{n_{ij}}_1\Delta^{n_{ij}}_2\Delta^{n_{ij}}_3\Delta^{n_{ij}}_4}
\end{equation}
where
\begin{equation}
P(x_i,\Delta_1,\Delta_2,\Delta_3,\Delta_4)=\frac{1}{(x_{12}^2)^{\frac{\Delta_1+\Delta_2}{2}}(x_{34}^2)^{\frac{\Delta_3+\Delta_4}{2}}}\left(\frac{x_{14}^2}{x_{24}^2}\right)^{\frac{\Delta_2-\Delta_1}{2}}\left(\frac{x_{14}^2}{x_{13}^2}\right)^{\frac{\Delta_3-\Delta_4}{2}}
\end{equation}
is the prefactor of $x_{ij}$ we extracted in (\ref{GUV}). From the above discussion we have already known how to decompose the $D$-function  $D_{\Delta^{n_{ij}}_1\Delta^{n_{ij}}_2\Delta^{n_{ij}}_3\Delta^{n_{ij}}_4}
$ into conformal blocks, when we view it as a four-point correlation function with external conformal dimensions $\Delta^{n_{ij}}_1$, $\Delta^{n_{ij}}_2$, $\Delta^{n_{ij}}_3$, $\Delta^{n_{ij}}_4$. The decomposition is given by (\ref{Dfunctioningxi}), but it will be helpful in the following to rewrite the decomposition in such a way that only the cross ratios are involved. Explicitly, we have
\begin{equation}\label{Dshifteddecomp}
\begin{split}
\frac{D_{\Delta^{n_{ij}}_1\Delta^{n_{ij}}_2\Delta^{n_{ij}}_3\Delta^{n_{ij}}_4}}{P(x_i,\Delta_1^{n_{ij}},\Delta_2^{n_{ij}},\Delta_3^{n_{ij}},\Delta_4^{n_{ij}})}={}&\sum_{n=0}^{\infty} a^{12,n_{ij}}_{n,0} g^{(s)}_{\Delta_1^{n_{ij}}+\Delta_2^{n_{ij}}+2n,0}(\epsilon,a^{n_{ij}},b^{n_{ij}};z,\bar{z})\\
{}&+\sum_{n=0}^{\infty} a^{34,n_{ij}}_{n,0} g^{(s)}_{\Delta_3^{n_{ij}}+\Delta_4^{n_{ij}}+2n,0}(\epsilon,a^{n_{ij}},b^{n_{ij}};z,\bar{z})\;
\end{split}
\end{equation}
where $a^{n_{ij}}=(\Delta_2^{n_{ij}}-\Delta_1^{n_{ij}})/2$ and $b^{n_{ij}}=(\Delta_3^{n_{ij}}-\Delta_4^{n_{ij}})/2$. In the above decomposition, we have restored all the parameters in a conformal block $g^{(s)}_{\Delta,\ell}(\epsilon,a,b;z,\bar{z})$ to emphasize its parameter dependence.  For a four-point correlator with external dimensions $\Delta_i$, the conformal block $g^{(s)}_{\Delta,\ell}(\epsilon,a,b;z,\bar{z})$ is a function of $\Delta$, $\ell$, the spacetime dimension, $z$, $\bar{z}$, and depends on the external conformal dimensions only via the combinations $a=(\Delta_2-\Delta_1)/2$, $b=(\Delta_3-\Delta_4)/2$. The latter fact has immediate interesting consequences. Consider for a moment the special example when $n_{13}=n_{14}=n_{24}=n_{23}=0$, it follows from (\ref{buildingblockandDshifted}) that we have the following identity
\begin{equation}
\frac{D^{\{n_{ij}\}}_{\Delta_1\Delta_2\Delta_3\Delta_4}}{P(x_i,\Delta_1,\Delta_2,\Delta_3,\Delta_4)}\bigg|_{n_{13}=n_{14}=n_{24}=n_{23}=0}=\frac{D_{\Delta^{n_{ij}}_1\Delta^{n_{ij}}_2\Delta^{n_{ij}}_3\Delta^{n_{ij}}_4}}{P(x_i,\Delta_1^{n_{ij}},\Delta_2^{n_{ij}},\Delta_3^{n_{ij}},\Delta_4^{n_{ij}})}\;.
\end{equation}
The conformal block decomposition (\ref{Dshifteddecomp}) of the RHS expresses the LHS as a sum of double-trace conformal blocks. To confirm that it really gives the conformal block decomposition for the LHS, we still need verify that the conformal blocks have the correct parameters, {\it i.e.}, $a^{n_{ij}}=a$, $b^{n_{ij}}=b$. We can easily check that this is indeed true. Therefore we conclude that when $n_{13}=n_{14}=n_{24}=n_{23}=0$, $D^{\{n_{ij}\}}_{\Delta_1\Delta_2\Delta_3\Delta_4}$ as a correlator with external dimensions $\Delta_k$, has the {\it same} decomposition as $D_{\Delta^{n_{ij}}_1\Delta^{n_{ij}}_2\Delta^{n_{ij}}_3\Delta^{n_{ij}}_4}$ as a correlator with dimensions $\Delta^{n_{ij}}_k$.

 More generally, we can relate the decomposition coefficients of a higher-derivative contact diagram to the those of a zero-derivative contact diagram using recursion relations of the conformal blocks. We can show that  (\ref{Dnijdef}) as a basis for decomposing $W^{bulk}$ is in fact {\it over complete}, and we can always restrict to $D^{\{n_{ij}\}}_{\Delta_1\Delta_2\Delta_3\Delta_4}$ with $n_{13}=n_{24}=n_{23}=n_{34}=0$ (we prove this in Appendix \ref{appcontact}). By inserting  (\ref{Dshifteddecomp}) into  (\ref{buildingblockandDshifted}) and setting $n_{13}=n_{24}=n_{23}=n_{34}=0$, we get 
\begin{equation}\label{Dnijdecomp}
\begin{split}
\frac{D^{\{n_{ij}\}}_{\Delta_1\Delta_2\Delta_3\Delta_4}}{P(x_i,\Delta_1,\Delta_2,\Delta_3,\Delta_4)}={}&(U^{-\frac{1}{2}})^{n_{14}} \sum_{n=0}^{\infty} a^{12,n_{ij}}_{n,0} g^{(s)}_{\Delta_1^{n_{ij}}+\Delta_2^{n_{ij}}+2n,0}(\epsilon,a^{n_{ij}},b^{n_{ij}};z,\bar{z})\\
+{}&(U^{-\frac{1}{2}})^{n_{14}}  \sum_{n=0}^{\infty} a^{34,n_{ij}}_{n,0} g^{(s)}_{\Delta_3^{n_{ij}}+\Delta_4^{n_{ij}}+2n,0}(\epsilon,a^{n_{ij}},b^{n_{ij}};z,\bar{z})\;.
\end{split}
\end{equation}
This is not a conformal block decomposition yet, due to the multiplicative factor $(U^{-\frac{1}{2}})^{n_{14}}$. To expand the RHS in conformal blocks, we use the following  recursion relation for conformal blocks\footnote{In this paper, we normalize our conformal blocks such that $g^{(s)}_{\Delta,\ell}\sim z^{\frac{\Delta-\ell}{2}}\bar{z}^{\frac{\Delta+\ell}{2}}$ when $z\ll\bar{z}\ll1$. This differs from the normalization used in \cite{Dolan:2011dv} by a factor: $\frac{(\epsilon)_\ell}{(2\epsilon)_\ell}g^{(s),{\bf Here}}_{\Delta,\ell}=g^{(s),{\bf DO}}_{\Delta,\ell}$, and $g^{(s),{\bf DO}}_{\Delta,\ell}$ was denoted as $F_{\lambda_1\lambda_2}$, with $\lambda_1=\frac{\Delta+\ell}{2}$, $\lambda_2=\frac{\Delta-\ell}{2}$.} \cite{Dolan:2011dv} (we suppress $\epsilon$ and $z$, $\bar{z}$ which are not changed)
\begin{equation}\label{Uhalfrecur}
\begin{split}
U^{-\frac{1}{2}}g^{(s)}_{\Delta,\ell}(a,b)={}&g^{(s)}_{\Delta-1,\ell}(a+\frac{1}{2},b+\frac{1}{2})+\Lambda_1\, g^{(s)}_{\Delta,\ell+1}(a+\frac{1}{2},b+\frac{1}{2})\\
{}&+\Lambda_2\, g^{(s)}_{\Delta,\ell-1}(a+\frac{1}{2},b+\frac{1}{2})+\Lambda_3\, g^{(s)}_{\Delta+1,\ell}(a+\frac{1}{2},b+\frac{1}{2})
\end{split}
\end{equation}
where
\begin{equation}
\begin{split}
\Lambda_1={}&-\frac{(2 a+\Delta +\ell ) (2 b+\Delta +\ell )}{4 (\Delta +\ell -1) (\Delta +\ell )}\;,\\
\Lambda_2={}&\frac{\ell  (\ell +2 \epsilon -1) (2 a+\Delta -\ell -2 \epsilon ) (-2 b-\Delta +\ell +2 \epsilon )}{4 (\ell +\epsilon -1) (\ell +\epsilon ) (-\Delta +\ell +2 \epsilon ) (-\Delta +\ell +2 \epsilon +1)}\;,\\
\Lambda_3={}&\frac{(\Delta -1) (\Delta -2 \epsilon ) (2 a+\Delta +\ell ) (2 b+\Delta +\ell ) (2 a+\Delta -\ell -2 \epsilon ) (2 b+\Delta -\ell -2 \epsilon )}{16 (\Delta -\epsilon -1) (\Delta -\epsilon ) (\Delta +\ell -1) (\Delta +\ell ) (-\Delta +\ell +2 \epsilon ) (-\Delta +\ell +2 \epsilon +1)}\;.
\end{split}
\end{equation}
After using the recursion relation for $n_{14}$ times, we can get rid of all powers of $U^{-\frac{1}{2}}$ in (\ref{Dnijdecomp}). Notice that the recursion coefficient $\Lambda_2$ contains a factor of $\ell$. This guarantees that when we apply this recursion relation to a spin-0 conformal block, we will not generate unphysical blocks with negative spins. Let us also note that each application of the recursion relation changes the values of $a$ and $b$ by $\frac{1}{2}$. Starting with $a^{n_{ij}}$, $b^{n_{ij}}$ and applying the recursion relation $n_{14}$ times, we can check that their values are shifted to
\begin{equation}
a^{n_{ij}}\to a^{n_{ij}}+\frac{n_{14}}{2}=\frac{\Delta_2-\Delta_1}{2}\;,\quad b^{n_{ij}}\to a^{n_{ij}}+\frac{n_{14}}{2}=\frac{\Delta_3-\Delta_4}{2}\;,
\end{equation}
matching precisely the $a$ and $b$ values that are assigned to a correlator with external dimensions $\Delta_1$, $\Delta_2$, $\Delta_3$, $\Delta_4$. 

To summarize,  we have the following algorithm for performing  conformal block decomposition of a generic contact Witten diagram $W^{contact}$
\begin{enumerate}
\item We write $W^{contact}$ as a linear combination of $D^{\{n_{ij}\}}_{\Delta_1\Delta_2\Delta_3\Delta_4}(x_i)$ with $n_{13}=n_{24}=n_{23}=n_{34}=0$.\footnote{It is easiest to find this linear combination in Mellin space (see Appendix \ref{appcontact}).}
\item Each $D^{\{n_{ij}\}}_{\Delta_1\Delta_2\Delta_3\Delta_4}(x_i)$ is associated with a $D$-function $D_{\Delta^{n_{ij}}_1\Delta^{n_{ij}}_2\Delta^{n_{ij}}_3\Delta^{n_{ij}}_4}$. We use (\ref{Dfunctioningxi}) to get the conformal block decomposition of $D_{\Delta^{n_{ij}}_1\Delta^{n_{ij}}_2\Delta^{n_{ij}}_3\Delta^{n_{ij}}_4}$ for all the $D^{\{n_{ij}\}}_{\Delta_1\Delta_2\Delta_3\Delta_4}$ that appear in the first step.
\item The conformal block decomposition of $D_{\Delta^{n_{ij}}_1\Delta^{n_{ij}}_2\Delta^{n_{ij}}_3\Delta^{n_{ij}}_4}$ in step 2 gives the RHS of (\ref{Dnijdecomp}). Using the relation (\ref{Uhalfrecur}) $n_{14}$ times, we obtain the conformal block decomposition of $D^{\{n_{ij}\}}_{\Delta_1\Delta_2\Delta_3\Delta_4}$.
\item We obtain the conformal block decomposition $W^{contact}$ by summing up the decompositions of $D^{\{n_{ij}\}}_{\Delta_1\Delta_2\Delta_3\Delta_4}$ with coefficients obtained from step 1.
\end{enumerate}

\subsection{Direct Channel Decomposition of Exchange Diagrams from Contact Diagrams}\label{Secdirectchan}
The direct channel decomposition of an exchange Witten diagram can be obtained by using the split representation (\ref{PiinOmega}) (\ref{Omegasplit}) for the bulk-to-bulk propagator, and this method has already been streamlined in \cite{Costa:2014kfa}. Schematically, because of (\ref{AdSlift}) and (\ref{Omegasplit}), each $\Omega^{\Delta',\ell}$ becomes proportional to a spin-$\ell$ conformal partial wave after integrations in position space. One obtains from this procedure the spectral representation of the exchange Witten diagram
\begin{equation}
W^{s,exchange}=\sum_{\ell=0}^{\ell_E}\int_{-i\infty}^{i\infty}\frac{dc}{2\pi i}\rho_\ell^{s,exchange}(c)\Psi^{(s)}_{\frac{d}{2}+c,\ell}(x_i)\;,
\end{equation}
which can further be written as 
\begin{equation}
W^{s,exchange}=\sum_{\ell=0}^{\ell_E}\int_{-i\infty}^{i\infty}\frac{dc}{2\pi i}\,2\rho_\ell^{s,exchange}(c)K^{\Delta_3,\Delta_3}_{\frac{d}{2}-c,\ell}g^{(s)}_{\frac{d}{2}+c,\ell}(x_i)\;,
\end{equation}
The OPE coefficients of the single-trace operator and double-trace operators can be obtained by closing the contour to the right and taking the residues.\footnote{For $\ell<J$, $\rho_\ell^{s,exchange}(c)$ also contains spurious poles which do not correspond to physical operators, see \cite{Costa:2012cb,Costa:2014kfa}.} We refer the reader to \cite{Costa:2014kfa} for details of implementing this algorithm.

The purpose of this subsection is to point out an alternative way of obtaining the direct channel decomposition, namely from the contact diagrams. We recall that the exchange Witten diagram satisfies  the following equation of motion identity (\ref{EOMWexWcon})
\begin{equation}\label{EOMWexWconb}
\left[\frac{1}{2}(\mathbf{L}_1+\mathbf{L}_2)^2+C^{(2)}_{\Delta_E,\ell_E}\right]W^{s,\, exchange}_{\Delta_E,\ell_E}=\sum_I c_IW^{contact}_I\;.
\end{equation}
In the previous subsection, we discussed how to decompose a generic contact Witten diagram into conformal blocks. Using the algorithm in Section \ref{SecCBdecompContact}, we can write the RHS of (\ref{EOMWexWconb}) as a sum of double-trace conformal blocks
\begin{equation}
\sum_I c_IW^{contact}_I=\sum_{J=0}^{\ell_E}\sum_{n=0}^\infty \tilde{a}^{12}_{n,J} g^{(s)}_{\Delta_1+\Delta_2+2n+J,J}(x_i)+\sum_{J=0}^{\ell_E}\sum_{n=0}^\infty \tilde{a}^{34}_{n,J} g^{(s)}_{\Delta_3+\Delta_4+2n+J,J}(x_i)\;.
\end{equation}
Because the double-trace conformal blocks are diagonal under the equation of motion operator with non vanishing eigenvalues, the double-trace coefficients of the exchange diagram are simply proportional to those of the contact Witten diagrams. This gives us right away all the double-trace coefficients $A^{12}_{n,J}$, $A^{34}_{n,J}$  in (\ref{Wexchangeins})
\begin{equation}\label{A12andA34}
A^{12}_{n,J}=\frac{\tilde{a}^{12}_{n,J}}{C^{(2)}_{\Delta_E,\ell_E}-C^{(2)}_{\Delta_1+\Delta_2+2n+J,J}}\;,\quad A^{34}_{n,J}=\frac{\tilde{a}^{34}_{n,J}}{C^{(2)}_{\Delta_E,\ell_E}-C^{(2)}_{\Delta_3+\Delta_4+2n+J,J}}\;.
\end{equation}
 We are therefore only one coefficient short for obtaining the full direct channel decomposition (\ref{Wexchangeins}). Naively, the information of the remaining single-trace block is lost because the single-trace block is annihilated by the equation of motion operator. However, knowing all these double-trace coefficients is in fact {\it enough} to uniquely fix the single-trace coefficient $A$. We can intuitively understand this statement from single-valuedness. As we alluded to in Section \ref{CPW}, a single conformal block is not single-valued in the Euclidean regime ($\bar{z}=z^*$), in contrast to an exchange Witten diagram. This fact implies that given we have obtained the correct double-trace coefficients $A^{12}_{n,J}$, $A^{34}_{n,J}$, there must be a unique choice for the single-trace coefficient $A$ such that the sum (\ref{Wexchangeins}) is single-valued.\footnote{One may also ask the following opposite question: with a fixed coefficient for the single-trace block, what should the double-trace coefficients be in order for the whole diagram to be single-valued? The answer to this question is not unique as one can always add contact diagrams which are single-valued.} More conceretely, we can relate the single-trace coefficient to the double-trace coefficients as follows. We start from the equation of motion identity (\ref{EOMWexWconb}). The algorithm in Section \ref{SecCBdecompContact} gives us the decomposition the contact diagrams on the RHS into double-trace operators, which can also be rewritten in terms of a spectral function
\begin{equation}
\sum_I c_IW^{contact}_I=\sum_{\ell=0}^{\ell_E}\int_{-i\infty}^{i\infty}\frac{dc}{2\pi i}\rho_\ell^{contact}(c)\Psi^{(s)}_{\frac{d}{2}+c,\ell}(x_i)\;.
\end{equation}
Notice that the conformal partial waves are diagonal under the equation of motion operator
\begin{equation}
\left[\frac{1}{2}(\mathbf{L}_1+\mathbf{L}_2)^2+C^{(2)}_{\Delta_E,\ell_E}\right]\Psi^{(s)}_{\frac{d}{2}+c,\ell}(x_i)=\left(C^{(2)}_{\Delta_E,\ell_E}-C^{(2)}_{\frac{d}{2}+c,\ell}\right)\Psi^{(s)}_{\frac{d}{2}+c,\ell}(x_i)\;.
\end{equation} 
This implies that the spectral function $\rho_\ell^{s,exchange}(c)$ of the exchange Witten diagram is related to $\rho_\ell^{contact}(c)$ by
\begin{equation}
\rho_\ell^{s,exchange}(c)=\frac{\rho_\ell^{contact}(c)}{C^{(2)}_{\Delta_E,\ell_E}-C^{(2)}_{\frac{d}{2}+c,\ell}}\;.
\end{equation}
In particular, the case with $\ell=\ell_E$
\begin{equation}
\rho_{\ell_E}^{s,exchange}(c)=\frac{\rho_{\ell_E}^{contact}(c)}{(\frac{d}{2}-\Delta_E)^2-c^2}\;,
\end{equation}
gives the coefficient for the spin-$\ell_E$ single-trace operator, upon taking the residue of the simple pole at $c=-\frac{d}{2}+\Delta_E$. 

\section{Recursion Relations in the Crossed Channel}\label{SecRecurcross}
\subsection{The Strategy for Crossed Channel Decomposition}\label{SecStrat}
In this subsection we outline a recursive method to compute the decomposition coefficients of exchange Witten diagrams and conformal partial waves in the crossed channel.\footnote{More precisely, we will consider the conformal block decomposition of a t-channel exchange Witten diagram in the s-channel. The decomposition of a u-channel exchange diagram in the s-channel is similar.} The technical details of this algorithm and further comments are left to the ensuing subsections. 

Our strategy exploits three useful facts.
 Firstly, it is well known that the exchange Witten diagrams can be decomposed into the crossed channel in terms of only double-trace conformal blocks. Secondly, as we  discussed in Section \ref{SecExtoCon}, acting on an exchange Witten diagram with the equation of motion operator, we can collapse it into contact Witten diagrams. The conformal block decomposition of the contact Witten diagrams is easy, and we have streamlined the method in Section \ref{SecCBdecompContact}. These two facts so far are not sufficient for making progress. We need to further combine them with the following third observation: the action of the t-channel equation of motion operator admits a very simple action on an s-channel conformal block. More precisely, the action produces a linear combination of finitely many conformal blocks with shifted quantum numbers. In particular, the double-trace spectra are preserved by the action of the equation of motion operator. Taken together, these three facts imply that the crossed channel decomposition coefficients of an exchange Witten diagram are not independent. Rather, the equation of motion operator reshuffles the  crossed channel double-trace operators of the exchange Witten diagram, and equates them to the decomposition of the contact Witten diagrams. The decomposition coefficients thus satisfy linear equations among themselves with the decomposition coefficients of contact diagrams as inhomogeneous terms. As we will soon see in detail, these linear equations and sufficiently simple thanks to the simple action of the equation of motion operator. They allow us to recursively solve all the crossed channel decomposition coefficients, when appropriate seed coefficients are inputted.  In one dimension, the seed coefficients are simply those of the lowest dimension operator in each double-trace tower, {\it i.e.}, the double-trace operators with dimension $\Delta_1+\Delta_2$
 and $\Delta_3+\Delta_4$.  In higher dimensions, the seed coefficients are those of the double-trace operators with minimal conformal twists, {\it i.e.}, $\tau=\Delta_1+\Delta_2$ and $\tau=\Delta_3+\Delta_4$. The same algorithm also applies to the crossed channel decomposition of conformal partial waves, since they also decompose into double-trace operators in the crossed channel and are annihilated by the equation of motion operator (therefore there are no inhomogeneous terms in the linear recursion relations).
 
In Section \ref{1d}, we flesh out the above schematic comments in the simplest case of $\mathrm{CFT}_1$. We derive the recursion relation for the crossed channel equation of motion operator acting on conformal blocks, and use these recursion relations to formulate a recursive algorithm. We discuss how to solve the crossed channel decomposition coefficients from the recursion relations. In particular we solve in a closed form the crossed channel decomposition of conformal partial waves, in terms of the Wilson polynomials. In Section \ref{higherd}, we detail the parallel story for $\mathrm{CFT}_d$ with $d>1$.

\subsection{One Dimension}\label{1d}
Let us consider an $AdS_2$ scalar exchange Witten diagram in the t-channel 
\begin{equation}
W^{t,exchange}_{\Delta_E}=\int \frac{d^{d+1}z_1}{z_{10}^{d+1}}\frac{d^{d+1}z_2}{z_{20}^{d+1}}G^{\Delta_1}_{B\partial}(z_1,x_1)G^{\Delta_4}_{B\partial}(z_1,x_4)\Pi^{\Delta_E}(z_1,z_2)G^{\Delta_2}_{B\partial}(z_2,x_2)G^{\Delta_3}_{B\partial}(z_2,x_3)\;.
\end{equation}
As we reviewed in Section \ref{SecExtoCon}, the diagram $W^{t,exchange}_{\Delta_E}$ satisfies the following t-channel equation of motion identity 
\begin{equation}\label{tchannelCasimir}
\left[\frac{1}{2}(\mathbf{L}_2+\mathbf{L}_3)^{AB}(\mathbf{L}_2+\mathbf{L}_3)_{AB}+M_E^2\right]W^{t,exchange}_{\Delta_E}=D_{\Delta_1\Delta_2\Delta_3\Delta_4}
\end{equation}
where $M_E^2=\Delta_E(\Delta_E-1)$. It is convenient to extract a kinematic factor in the convention of (\ref{GUV}), so that we work with functions of the cross ratio $z$
\begin{equation}
 G(x_i)=\frac{1}{(x_{12}^2)^{\frac{\Delta_1+\Delta_2}{2}}(x_{34}^2)^{\frac{\Delta_3+\Delta_4}{2}}}\left(\frac{x_{14}^2}{x_{24}^2}\right)^{a}\left(\frac{x_{14}^2}{x_{13}^2}\right)^{b}\mathcal{G}(z)\;.
\end{equation}
The equation of motion identity (\ref{tchannelCasimir}) then becomes
\begin{equation}\label{EOMtWexeqWcon}
\mathbf{EOM}^{(t)}[\mathcal{W}^{t,exchange}](z)=\mathcal{D}_{\Delta_1\Delta_2\Delta_3\Delta_4}(z)
\end{equation}
where $\mathcal{D}_{\Delta_1\Delta_2\Delta_3\Delta_4}(z)$ is $D_{\Delta_1\Delta_2\Delta_3\Delta_4}(x_i)$ after stripping off the kinematic factor. It is straightforward to work out the action of the differential operator $\mathbf{EOM}^{(t)}$ on a generic function $\mathcal{G}(z)$. The action takes the following form
\begin{equation}
\begin{split}
\mathbf{EOM}^{(t)}[\mathcal{G}]{}&=\frac{1}{2}\mathbf{D}_z(2a,2b)[\mathcal{G}(z)]-\mathbf{f}_0(2a,2b)[\mathbf{D}_z(2a,2b)[\mathcal{G}(z)]]+\Delta_E(\Delta_E-1)\mathcal{G}(z)\\
{}&-\left(1-\sum_{i=1}^4\Delta_i\right)\mathbf{f}_1(2a,2b)[\mathcal{G}(z)]-(\Delta_1+\Delta_2)(\Delta_3+\Delta_4)\mathbf{f}_0(2a,2b)[\mathcal{G}(z)]\\
{}&- \left(2ab+\frac{1}{2}\sum_{i=1}^4\Delta_i(\Delta_i-1)\right)\mathcal{G}(z)\;.
\end{split}
\end{equation}
In the above expression, we have additionally defined two operators
\begin{equation}
\mathbf{f}_0(a,b)=\frac{1}{z}-\frac{1}{2}\;, \quad \mathbf{f}_1(a,b)=(1-z)\frac{d}{dz}-\frac{1}{2}(a+b)\;.
\end{equation}

Let us now consider the action of $\mathbf{EOM}^{(t)}$ when $\mathcal{G}(z)$ is an s-channel conformal block. To proceed, the following properties of $\mathbf{f}_0$ and $\mathbf{f}_1$ \cite{Dolan:2011dv}  will be useful to us\footnote{Let us manifest the normalization by writing down the explicit expression for the 1d conformal block
\begin{equation}
g^{(s)}_\Delta(z)=z^\Delta{}_2F_1(\Delta+\Delta_2-\Delta_1,\Delta+\Delta_3-\Delta_4;2\Delta;z)\;.
\end{equation}}
\begin{equation}
\begin{split}
\mathbf{f}_0(a,b)[g_\Delta^{(s)}(z)]={}&g^{(s)}_{\Delta-1}(z)+\alpha_\Delta(a,b) g^{(s)}_{\Delta}(z)+\beta_\Delta(a,b) g^{(s)}_{\Delta+1}(z)\;,\\
\mathbf{f}_1(a,b)[g^{(s)}_\Delta(z)]={}&g^{(s)}_{\Delta-1}(z)+\alpha_\Delta(a,b) g^{(s)}_{\Delta}(z)-(\Delta-1)\beta_\Delta(a,b) g^{(s)}_{\Delta+1}(z)\;
\end{split}
\end{equation}
where 
\begin{equation}
\begin{split}
\alpha_\Delta(a,b)={}&-\frac{ab}{2\Delta(\Delta-1)}\;,\\
\beta_\Delta(a,b)={}&\frac{(\Delta+a)(\Delta+b)(\Delta-a)(\Delta-b)}{4\Delta^2(2\Delta-1)(2\Delta+1)}\;.
\end{split}
\end{equation}
Using these two relations and the Casimir equation
\begin{equation}
\mathbf{D}_z(2a,2b)[g^{(s)}_{\Delta}(z)]=\Delta(\Delta-1)g^{(s)}_{\Delta}(z)\;,
\end{equation}
we find the following three-term recursion relation for $\mathbf{EOM}^{(t)}$ acting on an s-channel conformal block
\begin{equation}\label{crosseomong}
\mathbf{EOM}^{(t)}[g^{(s)}_{\Delta}(z)]=\mu\,g^{(s)}_{\Delta-1}(z)+\nu\,g^{(s)}_{\Delta}(z)+\rho\,g^{(s)}_{\Delta+1}(z)\;.
\end{equation}
The recursion coefficients are given by
\begin{equation}
\begin{split}
\mu={}&-(\Delta -\Delta_1-\Delta_2) (\Delta -\Delta_3-\Delta_4)\;,\\
\nu={}&\frac{2 a b (\Delta_1+\Delta_2-1) (\Delta_3+\Delta_4-1)}{(\Delta -1) \Delta }+(\Delta_E -1) \Delta_E+\frac{1}{2} (\Delta -1) \Delta\\
{}&-\frac{1}{2} \left((\Delta_1-1)\Delta_1+(\Delta_2-1)\Delta_2+(\Delta_3-1)\Delta_3+(\Delta_4-1)\Delta_4\right)\;,\\
\rho={}&-\frac{(\Delta -2 a) (2 a+\Delta ) (\Delta -2 b) (2 b+\Delta ) (\Delta +\Delta_1+\Delta_2-1) (\Delta +\Delta_3+\Delta_4-1)}{4 \Delta ^2 (2 \Delta -1) (2 \Delta +1)}\;.
\end{split}
\end{equation}

Let us take the conformal dimension $\Delta$ to be the dimensions of the double-trace operators $\Delta_1+\Delta_2+n$. The recursion relation now reads
\begin{equation}\label{recurblock1d}
\mathbf{EOM}^{(t)}[g^{(s)}_{\Delta_1+\Delta_2+n}(z)]=\mu^{12}_n\,g^{(s)}_{\Delta_1+\Delta_2+n-1}(z)+\nu^{12}_n\,g^{(s)}_{\Delta_1+\Delta_2+n}(z)+\rho^{12}_n\,g^{(s)}_{\Delta_1+\Delta_2+n+1}(z)
\end{equation}
where $\mu^{12}_n$, $\nu^{12}_n$, $\rho^{12}_n$ are $\mu$, $\nu$, $\rho$ with $\Delta=\Delta_1+\Delta_2+n$. From the explicit expression of $\mu^{12}_n$ we can see that when $n=0$,
\begin{equation}
\mu^{12}_0=0\;,
\end{equation}
leaving only two double-trace blocks on the RHS of the recursion relation. The conformal block with dimension $\Delta_1+\Delta_2-1$ (which is not part of the double-trace spectrum) will not be generated. Therefore the action of $\mathbf{EOM}^{(t)}$ preserves the double-trace spectrum $\Delta_1+\Delta_2+n$ with $n\geq0$ and $n\in \mathbb{Z}$. A similar recursion relation also exists for double-trace operators with dimensions $\Delta_3+\Delta_4+n$, and can be obtained from the above relation by replacing $\Delta_1$, $\Delta_2$ with $\Delta_3$, $\Delta_4$.

These recursion relations give us an efficient way to compute the crossed channel decomposition. For simplicity, we will assume that the external conformal dimensions are generic such that no derivative conformal blocks appear. Special cases, such as $\Delta_i=\Delta_\phi$, can be obtained from the generic case by taking a limit, and will be discussed in Appendix \ref{appeqweight}.  Let us insert in (\ref{EOMtWexeqWcon}) the s-channel decomposition of the t-channel exchange diagram $\mathcal{W}^{t,exchange}$ 
\begin{equation}\label{texchange1dB12B34}
\mathcal{W}^{t,exchange}(z)=\sum_{n=0}^\infty B^{12}_n g^{(s)}_{\Delta_1+\Delta_2+n}(z)+\sum_{n=0}^\infty B^{34}_n g^{(s)}_{\Delta_3+\Delta_4+n}(z)\;.
\end{equation}
The relation (\ref{recurblock1d}) for $g^{(s)}_{\Delta_1+\Delta_2+n}$ and its counterpart for $g^{(s)}_{\Delta_3+\Delta_4+n}$ allows us to express the action of $\mathbf{EOM}^{(t)}$ on $\mathcal{W}^{t,exchange}(z)$ again in terms of the sum of double-trace conformal blocks. By the identity (\ref{EOMtWexeqWcon}), this decomposition should be equal to the decomposition (\ref{Dfunctioningxi}) for the contact diagram $\mathcal{D} _{\Delta_1\Delta_2\Delta_3\Delta_4}$\footnote{The  decomposition result (\ref{Dfunctioningxi}) also applies to the $AdS_2$ contact diagram. We just need to set $d=1$, and identify $g^{(s)}_{\Delta_1+\Delta_2+2n,0}$, $g^{(s)}_{\Delta_3+\Delta_4+2n,0}$ with the 1d conformal blocks $g^{(s)}_{\Delta_1+\Delta_2+2n}$, $g^{(s)}_{\Delta_3+\Delta_4+2n}$.} which takes the following form
\begin{equation}
\mathcal{D}_{\Delta_1\Delta_2\Delta_3\Delta_4}(z)=\sum_{n=0}^\infty a^{12}_n g^{(s)}_{\Delta_1+\Delta_2+2n}(z)+\sum_{n=0}^\infty a^{34}_n g^{(s)}_{\Delta_3+\Delta_4+2n}(z)\;.
\end{equation}
We therefore arrive at the following linear recursion relations among the crossed channel decomposition coefficients
\begin{equation}\label{recureq12inhomo}
\rho^{12}_{n-1}B^{12}_{n-1}+\nu^{12}_{n}B^{12}_{n}+\mu^{12}_{n+1}B^{12}_{n+1}=
\begin{cases}
a^{12}_{\frac{n}{2}}\;,\quad n\text{ even}\;,\\
0\;,\quad n\text{ odd}\;,
\end{cases}\quad \quad B^{12}_{-1}\equiv 0\;,
\end{equation}
\begin{equation}\label{recureq34inhomo}
\rho^{34}_{n-1}B^{34}_{n-1}+\nu^{34}_{n}B^{34}_{n}+\mu^{34}_{n+1}B^{34}_{n+1}=
\begin{cases}
a^{34}_{\frac{n}{2}}\;,\quad n\text{ even}\;,\\
0\;,\quad n\text{ odd}\;,
\end{cases}\quad \quad B^{34}_{-1}\equiv 0\;.
\end{equation}
It is straightforward to solve this relation. As we have commented already, when $n=0$ the three-term relations reduce to two-term relations with only $B^{12}_0$, $B^{12}_1$ and $B^{34}_0$, $B^{34}_1$. Once the seed coefficients $B^{12}_0$, $B^{34}_0$ are determined, the sub-leading coefficients $B^{12}_n$, $B^{34}_n$ with $n\geq 1$ can be recursively computed from the above three-term relations. In Appendix \ref{appseed}, we will show how to compute the seed coefficients.

\subsubsection*{Homogeneous Solution: the Wilson Polynomial}
The linear recursion equations (\ref{recureq12inhomo}) and (\ref{recureq34inhomo}) are inhomogeneous, and their solution gives the s-channel decomposition of a t-channel exchange Witten diagram. On the other hand, it is also interesting to consider the homogenous recursion equations for which the RHS' are  zero
\begin{equation}\label{recureq12homo}
\rho^{12}_{n-1}\widetilde{B}^{12}_{n-1}+\nu^{12}_{n}\widetilde{B}^{12}_{n}+\mu^{12}_{n+1}\widetilde{B}^{12}_{n+1}=0\;,
\end{equation}
\begin{equation}\label{recureq34homo}
\rho^{34}_{n-1}\widetilde{B}^{34}_{n-1}+\nu^{34}_{n}\widetilde{B}^{34}_{n}+\mu^{34}_{n+1}\widetilde{B}^{34}_{n+1}=0\;.
\end{equation}
 Such homogenous equations arise from a function $\mathcal{G}(z)$ that satisfies
\begin{equation}
\mathbf{EOM}^{(t)}[\mathcal{G}](z)=0\;,
\end{equation}
and admits an s-channel decomposition in terms of double-trace conformal blocks
\begin{equation}
\mathcal{G}(z)=\sum_{n=0}^\infty \widetilde{B}^{12}_n g^{(s)}_{\Delta_1+\Delta_2+n}(z)+\sum_{n=0}^\infty \widetilde{B}^{34}_n g^{(s)}_{\Delta_3+\Delta_4+n}(z)\;.
\end{equation}
In one dimension, such $\mathcal{G}(z)$ can be either the t-channel conformal block $g^{(t)}_{\Delta_E}(z)$, or the conformal partial wave $\Psi^{(t)}_{\Delta_E}(z)$. The solution to the homogenous recursion equations (\ref{recureq12homo}), (\ref{recureq34homo}) gives the crossed channel decomposition of the conformal block or partial wave (as we will see, their decomposition coefficients are only different by an overall factor). We will also see in the next subsection an extension of the story to $d>1$. There the solution to the homogenous recursion equations has to be interpreted as the crossed channel decomposition of a conformal partial wave. This is because Euclidean single-valuedness dictates that a single conformal block cannot be expressed in the crossed channel as infinitely many double-trace blocks, as we have already commented on at the end of Section \ref{CPW}.
 
Let us now solve the homogeneous equation (\ref{recureq12homo}). The solution to (\ref{recureq34homo}) can be obtained by simply replacing $\Delta_1$, $\Delta_2$ with $\Delta_3$, $\Delta_4$. Notice that a three-term recursion relation like (\ref{recureq12homo}) is characteristic of systems of orthogonal polynomials. Indeed, after making the change of variables 
\begin{equation}
\mathfrak{a}=\frac{1}{2}+\Delta_1-\Delta_4\;,\quad \mathfrak{b}=\frac{1}{2}+\Delta_2-\Delta_3\;,\quad \mathfrak{c}=-\frac{1}{2}+\Delta_1+\Delta_4\;,\quad \mathfrak{d}=-\frac{1}{2}+\Delta_2+\Delta_3\;, 
\end{equation}
and writing $\widetilde{B}^{12}_n$ as
\begin{equation}
\widetilde{B}^{12}_n=-\frac{\Gamma (\mathfrak{a}+\mathfrak{c}+n) \Gamma (\mathfrak{a}+\mathfrak{d}+n) \Gamma (\mathfrak{a}+\mathfrak{b}+\mathfrak{c}+\mathfrak{d}+n-1)}{\Gamma (n+1) \Gamma (\mathfrak{a}+\mathfrak{c}) \Gamma (\mathfrak{a}+\mathfrak{d}) \Gamma (\mathfrak{a}+\mathfrak{b}+\mathfrak{c}+\mathfrak{d}+2 n-1)}p_n\;,
\end{equation}
we find that the recursion relation (\ref{recureq12homo}) can be cast into the following form
\begin{equation}\label{Wilsonrecur}
-\left(\mathfrak{a}^2+x^2\right)p_n(x)=\mathcal{A}_n(p_{n+1}(x)-p_n(x))+\mathcal{B}_n(p_{n-1}(x)-p_n(x))
\end{equation}
where
\begin{equation}
x=i\left(\Delta_E-\frac{1}{2}\right)\;,
\end{equation}
\begin{equation}\label{Wilsonrecurcoe}
\begin{split}
\mathcal{A}_n={}&\frac{(\mathfrak{a}+\mathfrak{b}+n) (\mathfrak{a}+\mathfrak{c}+n) (\mathfrak{a}+\mathfrak{d}+n) (\mathfrak{a}+\mathfrak{b}+\mathfrak{c}+\mathfrak{d}+n-1)}{(\mathfrak{a}+\mathfrak{b}+\mathfrak{c}+\mathfrak{d}+2 n-1) (\mathfrak{a}+\mathfrak{b}+\mathfrak{c}+\mathfrak{d}+2 n)}\;,\\
\mathcal{B}_n={}&\frac{n (\mathfrak{b}+\mathfrak{c}+n-1) (\mathfrak{b}+\mathfrak{d}+n-1) (\mathfrak{c}+\mathfrak{d}+n-1)}{(\mathfrak{a}+\mathfrak{b}+\mathfrak{c}+\mathfrak{d}+2 n-2) (\mathfrak{a}+\mathfrak{b}+\mathfrak{c}+\mathfrak{d}+2 n-1)}\;.
\end{split}
\end{equation}
This is precisely the recursion relation that defines the Wilson polynomial \cite{Wilson1977,Wilson:1980aa}. Since the solution to the recursion relation is unique up to an overall rescaling, we can set $p_0(x)=1$ for convenience. Then $p_n(x)$ can be expressed compactly as a ${}_4F_3$ function \cite{Wilson1977,Wilson:1980aa}
\begin{equation}\label{Wilsonpolyn}
p_n(x;\mathfrak{a},\mathfrak{b},\mathfrak{c},\mathfrak{d})={}_4F_3\left(\left.\begin{array}{c}-n,n+\mathfrak{a}+\mathfrak{b}+\mathfrak{c}+\mathfrak{d}-1, \mathfrak{a}+ix,\mathfrak{a}-ix\\\mathfrak{a}+\mathfrak{b},\mathfrak{a}+\mathfrak{c},\mathfrak{a}+\mathfrak{d}\end{array}\right.;1\right)\;.
\end{equation}
This function $p_n(x)$ is a polynomial in $x^2$ of degree $n$, as is already clear from the recursion relation (\ref{Wilsonrecur}). 

Now let us use this solution in the s-channel decomposition of a 1d t-channel conformal block. The decomposition takes the following form
\begin{equation}\label{tconfblockdecomp}
\begin{split}
\frac{z^{\Delta_1+\Delta_2}}{(1-z)^{\Delta_2+\Delta_3}}g^{(t)}_{\Delta_E}(z)={}&\frac{z^{\Delta_1+\Delta_2}}{(1-z)^{\Delta_2+\Delta_3}}{}_2F_1(\Delta_E-\Delta_3+\Delta_2,\Delta_E+\Delta_1-\Delta_4,2\Delta_E,1-z)\\
={}&\sum_{n=0}^{\infty}b^{12}_n g^{(s)}_{\Delta_1+\Delta_2+n}(z)+\sum_{n=0}^{\infty}b^{34}_n g^{(s)}_{\Delta_3+\Delta_4+n}(z)\;,
\end{split}
\end{equation}
and one can check that the LHS of (\ref{tconfblockdecomp}) is annihilated by the equation of motion operator $\mathbf{EOM}^{(t)}$. The action on the RHS thus gives rise to the recursion equations (\ref{recureq12homo}) and (\ref{recureq34homo}). The above solution to the homogenous equations (\ref{recureq12homo}) then determines the ratios $b^{12}_n/b^{12}_0$ of the decomposition coefficients to be
\begin{equation}\label{b12ratio}
\frac{b^{12}_n}{b^{12}_0}=\frac{\Gamma (n+1) (\mathfrak{a}+\mathfrak{b}+\mathfrak{c}+\mathfrak{d}+n-1)_n}{(\mathfrak{a}+\mathfrak{c})_n (\mathfrak{a}+\mathfrak{d})_n}p_n(x;\mathfrak{a},\mathfrak{b},\mathfrak{c},\mathfrak{d})\;.
\end{equation}
By further comparing the two sides of (\ref{tconfblockdecomp}) expanded at $z=1^-$, we can also determine the leading coefficient
\begin{equation}
b^{12}_0=\frac{\Gamma (2 \Delta_E) \Gamma (-\Delta_1-\Delta_2+\Delta_3+\Delta_4)}{\Gamma (-\Delta_1+\Delta_4+\Delta_E) \Gamma (-\Delta_2+\Delta_3+\Delta_E)}\;,
\end{equation}
and thereby obtaining the crossed channel decomposition. Similarly, $b^{34}_n$ can also be obtained by solving recursion relations. Its solution is simply the solution for $b^{12}_n$ with $\Delta_1$, $\Delta_2$ replaced by $\Delta_3$, $\Delta_4$. The above crossed channel decomposition coefficients can also be obtained from the alpha space techniques \cite{Hogervorst:2017sfd}.

Note that the coefficient ratio (\ref{b12ratio}) depends on $\Delta_E$ only via the shadow symmetric combination $x^2$. This implies that the decomposition coefficients for the shadow conformal block have the same ratio (\ref{b12ratio}). Furthermore, since the conformal partial wave (\ref{CPWdef}) is just the linear combination of the conformal block and its shadow, its crossed channel decomposition coefficients will also have the same ratio.

\subsection{$d>1$ Dimensions}\label{higherd}
The story for $d>1$ is similar to what we have seen for $d=1$. We will organize this subsection as the follows.  We start by obtaining the action of the t-channel equation of motion operator on a generic conformal block. We will find the result can be expressed as the linear combination of five conformal blocks with shifted dimensions and spins. Then we restrict the quantum numbers of the conformal blocks to those of the double-trace operators which appear in the crossed channel decomposition. The spectra of these operators are preserved by the recursion relation. This gives us an efficient algorithm for recursively solving the crossed channel decomposition coefficients. We can consider two different problems depending on whether the recursion equations are homogenous or inhomogeneous. The latter corresponds to the s-channel decomposition of a t-channel exchange Witten diagram, while the former corresponds to the decomposition of an t-channel conformal partial wave. We also consider the special case of equal external weights, where our results imply recursion relations for the anomalous dimensions.

\subsubsection*{Action of the t-channel equation of motion operator}
Consider a t-channel exchange Witten diagram with dimension $\Delta_E$ and spin $\ell_E$. The t-channel equation of motion give the identity 
\begin{equation}\label{higherdtchEOM}
\left(\frac{1}{2}(\mathbf{L}_2+\mathbf{L}_3)^2+C^{(2)}_{\Delta_E,\ell_E}\right)W^{t,exchange}_{\Delta_E,\ell_E}=\sum_I c_I W^{contact}_I\;.
\end{equation}
The LHS of the identity defines a second order differential operator $\mathbf{EOM}^{(t)}$ which acts on $\mathcal{G}(z,\bar{z})$
\begin{equation}\small
\begin{split}
{}&\mathbf{EOM}^{(t)}[\mathcal{G}(z,\bar{z})]=\mathbf{\Delta}_\epsilon(a,b)[\mathcal{G}(z,\bar{z})]-\mathbf{F}_0(a,b)[\mathbf{\Delta}_\epsilon(a,b)[\mathcal{G}(z,\bar{z})]]+M_E^2\, \mathcal{G}(z,\bar{z})\\
{}&+\mathbf{F}_2(a,b)[\mathcal{G}(z,\bar{z})]+\left(\Delta_1+\Delta_2+\Delta_3+\Delta_4-2+2\epsilon\right)\mathbf{F}_1(a,b)[\mathcal{G}(z,\bar{z})]\\
{}&-\frac{1}{2}(\Delta_1+\Delta_2)(\Delta_3+\Delta_4)\mathbf{F}_0(a,b)[\mathcal{G}(z,\bar{z})]\\
{}&+\left(2+\epsilon(\Delta_1+\Delta_2+\Delta_3+\Delta_4)-\frac{(\Delta_1-1)^2+(\Delta_2-1)^2+(\Delta_3-1)^2+(\Delta_4-1)^2}{2}-2ab\right)\mathcal{G}(z,\bar{z})
\end{split}
\end{equation}
where $\mathbf{F}_i$ is a differential operator of order $i$
\begin{equation}
\begin{split}
\mathbf{F}_0(a,b)={}&\frac{1}{z}+\frac{1}{\bar{z}}-1\;,\\
\mathbf{F}_1(a,b)={}&(1-z)\frac{\partial}{\partial z}+(1-\bar{z})\frac{\partial}{\partial \bar{z}}\;,\\
\mathbf{F}_2(a,b)={}&\frac{z-\bar{z}}{z\bar{z}}(\mathbf{D}_z(a,b)-\mathbf{D}_{\bar{z}}(a,b))\;.
\end{split}
\end{equation}
Here it will turn out to be more convenient to use the normalization of \cite{Dolan:2011dv} for the conformal blocks, which are denoted by $F_{\lambda_1,\lambda_2}(z,\bar{z})$ 
\begin{equation}
\begin{split}
F_{\lambda_1,\lambda_2}(z,\bar{z})={}&\mathcal{N}_{\epsilon,\ell}g^{(s)}_{\Delta,\ell}(z,\bar{z})\;,\\
\mathcal{N}_{\epsilon,\ell}={}&\frac{(\epsilon)_\ell}{(2\epsilon)_\ell}\;,\\
\lambda_1=\frac{\Delta+\ell}{2}\;,{}&\quad \lambda_2=\frac{\Delta-\ell}{2}\;.
\end{split}
\end{equation}
The $\mathbf{F}_i$ operators have the following recursion relations on conformal blocks \cite{Dolan:2011dv}
\begin{equation}\label{Ffivetermrecur}
\mathbf{F}_i(a,b)[\mathcal{F}_{\lambda_1,\lambda_2}]=r_i F_{\lambda_1,\lambda_2-1}+s_iF_{\lambda_1-1,\lambda_2}+t_iF_{\lambda_1+1,\lambda_2}+u_iF_{\lambda_1,\lambda_2+1}+w_iF_{\lambda_1,\lambda_2}\;,
\end{equation}
with various coefficients defined as follows. The coefficients with label $i=0$ are
\begin{equation}
\begin{split}
{}&r_0=\frac{\lambda_1-\lambda_2+2 \epsilon }{\lambda_1-\lambda_2+\epsilon }\;,\quad s_0=\frac{\lambda_1-\lambda_2}{\lambda_1-\lambda_2+\epsilon }\;,\\
{}&t_0=\frac{(\lambda_1+\lambda_2-1) (\lambda_1+\lambda_2-2 \epsilon ) (\lambda_1-\lambda_2+2 \epsilon )}{(\lambda_1+\lambda_2-\epsilon -1) (\lambda_1+\lambda_2-\epsilon ) (\lambda_1-\lambda_2+\epsilon )}\beta_{\lambda_1}(a,b)\;,\\
{}&u_0=\frac{(\lambda_1-\lambda_2) (\lambda_1+\lambda_2-1) (\lambda_1+\lambda_2-2 \epsilon )}{(\lambda_1+\lambda_2-\epsilon -1) (\lambda_1+\lambda_2-\epsilon ) (\lambda_1-\lambda_2+\epsilon )}\beta_{\lambda_2-\epsilon}(a,b)\;,\\
{}&w_0=-\frac{(c_{\lambda_1\lambda_2}+2\epsilon)}{2\lambda_1(\lambda_1-1)(\lambda_2-\epsilon)(\lambda_2-1-\epsilon)} ab\;.
\end{split}
\end{equation}
In terms of the $i=0$ coefficients, the higher $i$ coefficients are given by\footnote{There are two typos in (4.29) of \cite{Dolan:2011dv}. The first typo is in the expression for $s_2$ and is corrected below. The other one is in $s_3$, and the correct expression should be
\begin{equation}
s_3=(\lambda_1+\epsilon)(\lambda_1-\lambda_2+2\epsilon)(\lambda_1+\lambda_2-1)s_0\;.
\end{equation}}
\begin{equation}
\begin{split}
{}&r_1=\lambda_2 r_0\;,\quad s_1=(\lambda_1+\epsilon)s_0\;,\quad t_1=-(\lambda_1-1-\epsilon)t_0\;,\quad u_1=-(\lambda_2-1-2\epsilon)u_0\;,\\
{}& w_1=(1+\epsilon)w_0\;,
\end{split}
\end{equation}
\begin{equation}
\begin{split}
{}& r_2=(\lambda_1-\lambda_2) (\lambda_1+\lambda_2-1)r_0\;,\quad s_2= -(\lambda_1+\lambda_2-1) (\lambda_1-\lambda_2+2\epsilon )s_0\;,\\
{}& t_2=-(\lambda_1-\lambda_2) (\lambda_1+\lambda_2-2 \epsilon -1)t_0\;,\quad u_2=(\lambda_1-\lambda_2+2 \epsilon ) (\lambda_1+\lambda_2-2 \epsilon -1)u_0\;,\\
{}& w_2=-\frac{(\lambda_1-\lambda_2) (\lambda_1+\lambda_2-1) (\lambda_1-\lambda_2+2 \epsilon ) (\lambda_1+\lambda_2-2 \epsilon -1)}{2\lambda_1 (\lambda_1-1) (\lambda_2-\epsilon ) (\lambda_2-\epsilon -1)}ab\;.
\end{split}
\end{equation}
Using the recursion relations (\ref{Ffivetermrecur}) and the Casimir equation
\begin{equation}
2\mathbf{\Delta}_\epsilon(a,b)[F_{\lambda_1,\lambda_2}]=C^{(2)}_{\Delta,\ell}F_{\lambda_1,\lambda_2}\;,
\end{equation}
 the action of $\mathbf{EOM}^{(t)}$ on a conformal block $F_{\lambda_1,\lambda_2}$ can be expressed as the linear combination of five conformal blocks with shifted dimensions and spins
\begin{equation}\label{EOMrecuranyd}
\mathbf{EOM}^{(t)}[\mathcal{F}_{\lambda_1,\lambda_2}]=\mathfrak{R}\, F_{\lambda_1,\lambda_2-1}+\mathfrak{S}\,F_{\lambda_1-1,\lambda_2}+\mathfrak{T}\,F_{\lambda_1+1,\lambda_2}+\mathfrak{U}\,F_{\lambda_1,\lambda_2+1}+\mathfrak{W}\,F_{\lambda_1,\lambda_2}\;.
\end{equation}
The coefficients are determined to be\footnote{Here we can appreciate the advantage of using the normalization of \cite{Dolan:2011dv}: the coefficients $\mathfrak{S}$ and $\mathfrak{U}$ are simply related to $\mathfrak{R}$ and $\mathfrak{T}$ with $\ell\to-\ell-2\epsilon$. Otherwise there will be an additional factor which depends on $\epsilon$ and $\ell$. }
\begin{equation}\label{EOMrecuranydcoe}
\begin{split}
\mathfrak{R}={}&-\frac{(\ell +2 \epsilon ) (-\Delta +\Delta_1+\Delta_2+\ell ) (-\Delta +\Delta_3+\Delta_4+\ell )}{2 (\ell +\epsilon )}\,\\
\mathfrak{T}={}&-\frac{(2 a+\Delta +\ell ) (-2 a+\Delta +\ell ) (2 b+\Delta +\ell ) (-2 b+\Delta +\ell )}{32 (\Delta -\epsilon )  (\Delta +\ell )^2 (\Delta -\epsilon -1) }\\
{}&\times \frac{(\Delta +\Delta_1+\Delta_2+\ell -2 \epsilon -2) (\Delta +\Delta_3+\Delta_4+\ell -2 \epsilon -2)(\Delta -1) (\Delta -2 \epsilon )(\ell +2 \epsilon)}{(\Delta +\ell -1)(\Delta +\ell +1) (\ell +\epsilon )}\,\\
\mathfrak{S}={}&\mathfrak{R}\big|_{\ell\to-\ell-2\epsilon}\;,\\
\mathfrak{U}={}&\mathfrak{T}\big|_{\ell\to-\ell-2\epsilon}\;.\\
\mathfrak{W}={}&C^{(2)}_{\Delta_E,\ell_E}+\frac{C^{(2)}_{\Delta,\ell}-\sum_{i=1}^4C^{(2)}_{\Delta_i,\ell_i=0}}{2}+\frac{2ab(\Delta_1+\Delta_2-2\epsilon-2)(\Delta_3+\Delta_4-2\epsilon-2)(C^{(2)}_{\Delta,\ell}+4\epsilon)}{(\Delta +\ell -2) (\Delta +\ell ) (-\Delta +\ell +2 \epsilon ) (-\Delta +\ell +2 \epsilon +2)}
\end{split}
\end{equation}
Note that the above recursion relation has the following desirable features similar to those of (\ref{crosseomong}). We first look at the factor $\mathfrak{R}$ multiplying the conformal block $F_{\lambda_1,\lambda_2-1}$ which has conformal dimension $\Delta-1$ and twist $\Delta-\ell$. This factor vanishes $\mathfrak{R}$ when the conformal block $F_{\lambda_1,\lambda_2}$ has the minimal twist for a double-trace operator formed with $\mathcal{O}_1$ and $\mathcal{O}_2$ or with $\mathcal{O}_3$ and $\mathcal{O}_4$, {\it i.e.}, 
\begin{equation}
\mathfrak{R}=0\;,\quad\text{when}\quad\Delta=\ell+\Delta_1+\Delta_2\;,\quad \text{or}\quad \Delta=\ell+\Delta_3+\Delta_4\;.
\end{equation}
Let us also notice that the coefficients $\mathfrak{S}$ and $\mathfrak{U}$, which multiply conformal blocks with shifted spin $\ell-1$, both  contain a factor $\ell$. It implies that when the spin of $F_{\lambda_1,\lambda_2}$ is zero, both $\mathfrak{S}$ and $\mathfrak{U}$ vanish and no blocks with negative spins will be generated on the RHS. Moreover, the twists of conformal blocks are always shifted by an even integer. It is therefore not hard to see that these properties of the coefficients guarantee that the double-trace spectra, labelled by $\{\Delta,\ell\}$ with
\begin{equation}
\Delta=\Delta_1+\Delta_2+2n+\ell\quad \text{and}\quad \Delta=\Delta_3+\Delta_4+2n+\ell\;,\quad n,\ell=0,1,2,\ldots,
\end{equation}
are preserved by the recursion relation (\ref{EOMrecuranyd}). 

\subsubsection*{Reduction to 1d}
We can show that the recursion relation in $d>1$ reduces to the 1d recursion relation (\ref{crosseomong}) in an appropriate limit. Setting $\epsilon=-\frac{1}{2}$, $\ell_E=0$, $\ell=0$ and restricting to $z=\bar{z}$ we find from  (\ref{EOMrecuranydcoe}) that
\begin{equation}
\mathfrak{S}=\mathfrak{U}=0\;,\quad \mathfrak{R}=\mu\;,\quad \mathfrak{W}=\nu\;,\quad \mathfrak{T}=\rho\;.
\end{equation}
We then use the following identity for conformal blocks\footnote{In the second identity, $\ell=1$ for the conformal block $F_{\lambda_1\lambda_2}$. Note that $\ell$ appears in the Casimir eigenvalue as $\ell(\ell+2\epsilon)=\ell(\ell-1)$, both $\ell=0$ and $\ell=1$ give the same 1d conformal block.} \cite{Dolan:2011dv}
\begin{equation}
F^{(-\frac{1}{2})}_{\frac{\Delta}{2}\frac{\Delta}{2}}(z,z)=g_\Delta(z)\;,\quad F^{(-\frac{1}{2})}_{\frac{\Delta+1}{2}\frac{\Delta-1}{2}}(z,z)=g_\Delta(z)\;,
\end{equation}
to reduce the higher dimensional conformal blocks to one dimensional blocks $g_\Delta(z)$. We find that the 1d recursion relation (\ref{crosseomong}) is precisely reproduced.

\subsubsection*{The recursive algorithm for the crossed channel decomposition}
Let us now use the recursion relation (\ref{EOMrecuranyd}) to formulate an algorithm for the crossed channel decomposition of an exchange Witten diagram.  We start with the s-channel decomposition for $W^{t,\,exchange}_{\Delta_E,\ell_E}$, with the assumption that the external dimensions $\Delta_i$ are generic
\begin{equation}
W^{t,\,exchange}_{\Delta_E,\ell_E}=\sum_{J=0}^{\infty}\sum_{n=0}^\infty \underbrace{\mathcal{B}^{12}_{n,J} \mathcal{N}_{\epsilon,J}}_{B^{12}_{n,J}} g^{(s)}_{\Delta_1+\Delta_2+2n+J,J}(x_i)+\sum_{J=0}^{\infty}\sum_{n=0}^\infty \underbrace{\mathcal{B}^{34}_{n,J}\mathcal{N}_{\epsilon,J}}_{B^{34}_{n,J}} g^{(s)}_{\Delta_3+\Delta_4+2n+J,J}(x_i)\;.
\end{equation}
Here $\mathcal{N}_{\epsilon,J}=\frac{(\epsilon)_J}{(2\epsilon)_J}$ is a normalization factor such that $\mathcal{N}_{\epsilon,J} g^{(s)}_{\Delta,J}=F_{\frac{\Delta+J}{2},\frac{\Delta-J}{2}}$. We insert this decomposition in (\ref{higherdtchEOM}). Thanks to the recursion relation (\ref{EOMrecuranyd}), the action of $\mathbf{EOM}^{(t)}$ can be rewritten as the linear combination of double-trace blocks. The resulting expansion should be equal to the conformal block decomposition of the contact terms on the RHS
\begin{equation}
\sum_I c_I W_I^{contact}=\sum_{J=0}^{\ell_E}\sum_{n=0}^\infty \tilde{a}^{12}_{n,J} g^{(s)}_{\Delta_1+\Delta_2+2n+J,J}(x_i)+\sum_{J=0}^{\ell_E}\sum_{n=0}^\infty \tilde{a}^{34}_{n,J} g^{(s)}_{\Delta_3+\Delta_4+2n+J,J}(x_i)\;.
\end{equation}
For convenience, let us define $\mathfrak{R}^{12}_{n,\ell}$, $\mathfrak{S}^{12}_{n,\ell}$, $\mathfrak{T}^{12}_{n,\ell}$, $\mathfrak{U}^{12}_{n,\ell}$, $\mathfrak{W}^{12}_{n,\ell}$ to be $\mathfrak{R}$, $\mathfrak{S}$, $\mathfrak{T}$, $\mathfrak{U}$, $\mathfrak{W}$ with $\Delta=\Delta_1+\Delta_2+2n+\ell$, and  $\mathfrak{R}^{34}_{n,\ell}$, $\mathfrak{S}^{34}_{n,\ell}$, $\mathfrak{T}^{34}_{n,\ell}$, $\mathfrak{U}^{34}_{n,\ell}$, $\mathfrak{W}^{34}_{n,\ell}$ to be $\mathfrak{R}$, $\mathfrak{S}$, $\mathfrak{T}$, $\mathfrak{U}$, $\mathfrak{W}$ with $\Delta=\Delta_3+\Delta_4+2n+\ell$.
Then we arrive at the following recursion relations for the crossed channel decomposition coefficients. The recursion relation for $\mathcal{B}^{12}_{n,\ell}$ is
\begin{equation}\label{Recur12nleqatilde12nl}
\begin{split}
Recur^{12}_{n,\ell}\equiv\mathfrak{R}^{12}_{n+1,\ell-1}\mathcal{B}^{12}_{n+1,\ell-1}+\mathfrak{S}^{12}_{n,\ell+1}\mathcal{B}^{12}_{n,\ell+1}+\mathfrak{T}^{12}_{n,\ell-1}\mathcal{B}^{12}_{n,\ell-1}{}&\\
+\mathfrak{U}^{12}_{n-1,\ell+1}\mathcal{B}^{12}_{n-1,\ell+1}+\mathfrak{W}^{12}_{n,\ell}\mathcal{B}^{12}_{n,\ell}{}&=\tilde{a}^{12}_{n,\ell}\;,
\end{split}
\end{equation}
with $n\geq0$, $\ell\geq0$, and 
\begin{equation}
\mathcal{B}^{12}_{n,-1}=\mathcal{B}^{12}_{-1,\ell}=0\;.
\end{equation}
The recursion relation for $\mathcal{B}^{34}_{n,\ell}$ takes the same form, and can be obtained by replacing 1, 2 with 3, 4.

\begin{figure}[t]
\begin{center}
\includegraphics[scale=0.68]{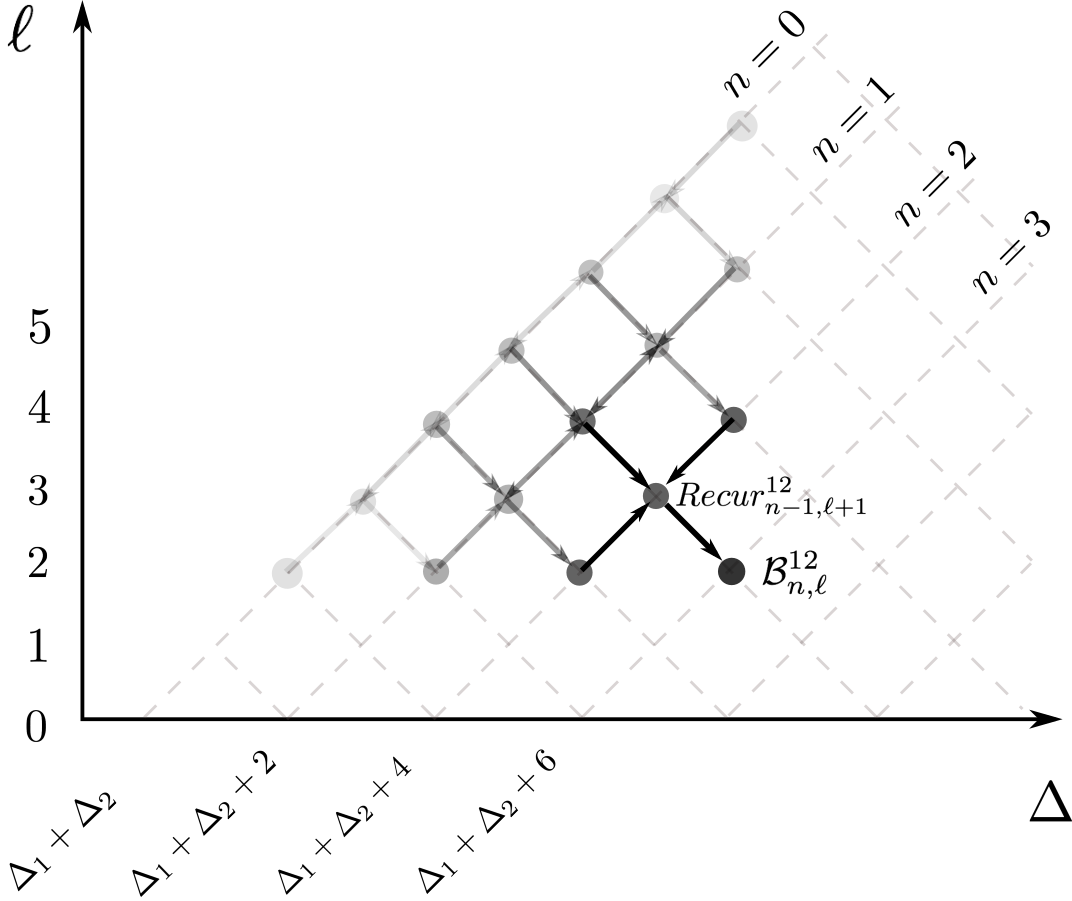}
\caption{Illustration of the recursive algorithm for solving the OPE coefficient $\mathcal{B}^{12}_{n,\ell}$. The equation $Recur^{12}_{n-1,\ell+1}=\tilde{a}^{12}_{n-1,\ell+1}$ solves $\mathcal{B}^{12}_{n,\ell}$ in terms of $\mathcal{B}^{12}_{n-1,\ell}$, $\mathcal{B}^{12}_{n-1,\ell+1}$, $\mathcal{B}^{12}_{n-1,\ell+2}$ and $\mathcal{B}^{12}_{n-2,\ell+2}$ which have smaller values of $n$. This procedure is iterated until one reaches the $n=0$ data which are the seed coefficients one inputs.}
\label{recuralgo}
\end{center}
\end{figure}

Let us discuss how we can solve these recursion relations. Without the loss of generality, we focus on the coefficients $\mathcal{B}^{12}_{n,\ell}$ and their recursion relations. Just as in the 1d case where we need to input the seed coefficient for the double-trace operator with minimal conformal dimension, here we need to input the OPE coefficients of the double-trace operators with minimal {\it twist}, {\it i.e.}, $\mathcal{B}^{12}_{0,\ell}$. These seed coefficients can be obtained from applying the inversion formula \cite{Caron-Huot:2017vep} for $\ell>\ell_E$ \cite{Liu:2018jhs,Cardona:2018dov}, or from Mellin space \cite{Costa:2014kfa,Sleight:2018epi,Sleight:2018ryu}.  After inputting the seed coefficients, the coefficients $\mathcal{B}^{12}_{n,\ell}$ of operators with higher twists are uniquely determined. More precisely,  the equation 
\begin{equation}\label{eqn12nllgeq1}
Recur^{12}_{n-1,\ell+1}=\tilde{a}^{12}_{n-1,\ell+1}\;,\quad     n\geq1\;, \ell\geq0\;,
\end{equation}
determines the coefficient $\mathcal{B}^{12}_{n,\ell}$ in terms of $\mathcal{B}^{12}_{n',\ell}$ with $n'<n$ 
\begin{equation}\label{solB12nl}
\mathcal{B}^{12}_{n,\ell}=\frac{\tilde{a}^{12}_{n-1,\ell+1}-\mathfrak{S}^{12}_{n-1,\ell+2}\mathcal{B}^{12}_{n-1,\ell+2}-\mathfrak{T}^{12}_{n-1,\ell}\mathcal{B}^{12}_{n-1,\ell}-\mathfrak{U}^{12}_{n-2,\ell+2}\mathcal{B}^{12}_{n-2,\ell+2}-\mathfrak{W}^{12}_{n-1,\ell+1}\mathcal{B}^{12}_{n-1,\ell+1}}{\mathfrak{R}^{12}_{n,\ell}}\;.
\end{equation}
We can start with $Recur^{12}_{0,\ell+1}=\tilde{a}^{12}_{0,\ell+1}$ to solve $\mathcal{B}^{12}_{1,\ell}$ in terms of $\mathcal{B}^{12}_{0,\ell}$ (note $\mathcal{B}^{12}_{-1,\ell}=0$), and work our way up to any $\mathcal{B}^{12}_{n,\ell}$ by considering equations $Recur^{12}_{i,\ell+1}=\tilde{a}^{12}_{i,\ell+1}$ with increasing $i$. This solving strategy is pictorially illustrated by Figure \ref{recuralgo}.

\begin{figure}[t]
\begin{center}
\includegraphics[scale=0.68]{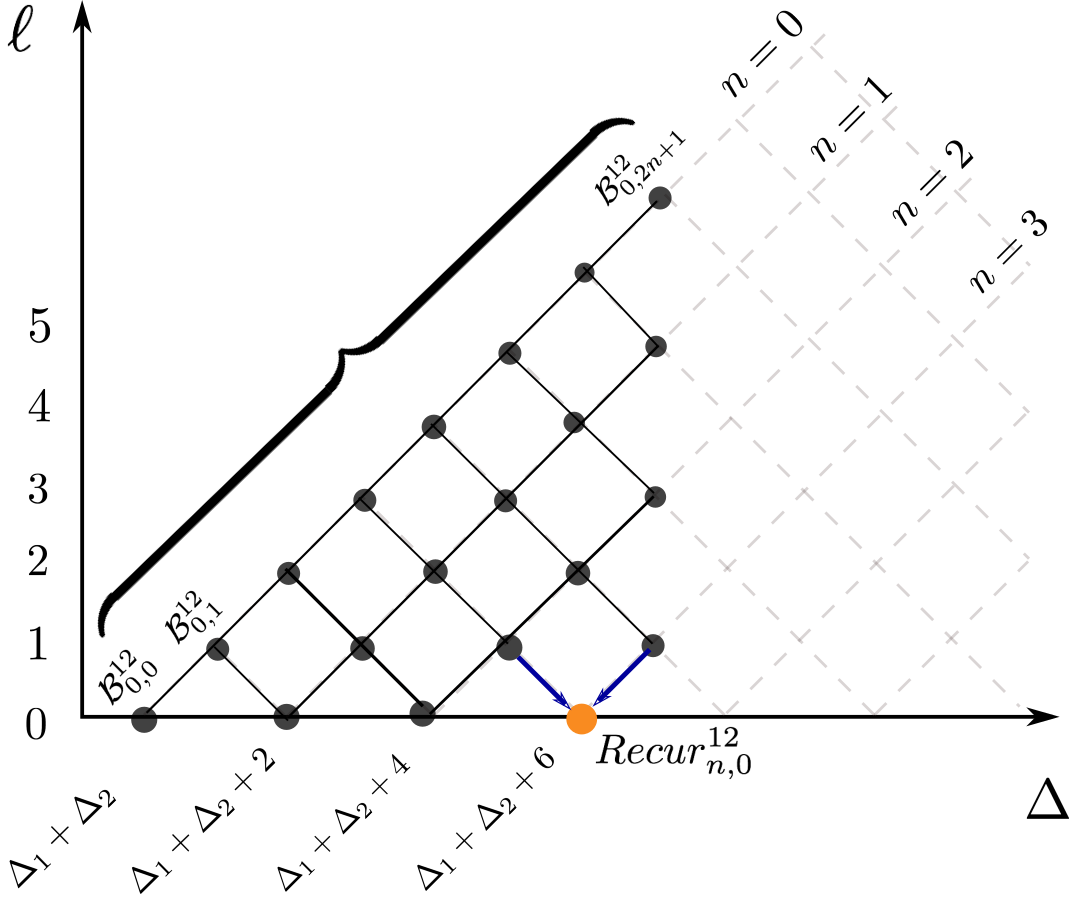}
\caption{Illustration of the constraints on $\mathcal{B}^{12}_{0,0}$, $\mathcal{B}^{12}_{0,1}$, \ldots, $\mathcal{B}^{12}_{0,2n+1}$ imposed by the equation $Recur^{12}_{n,0}=\tilde{a}^{12}_{n,0}$.}
\label{recurconstrs}
\end{center}
\end{figure}

Although the recursion relations (\ref{Recur12nleqatilde12nl}) are not enough to fix the leading twist OPE coefficients $\mathcal{B}^{12}_{0,\ell}$, we should note that they still impose nontrivial constraints on the possible values that $\mathcal{B}^{12}_{0,\ell}$ can take as a function of $\ell$. We note that the equations (\ref{eqn12nllgeq1}) have not exhausted all the cases of (\ref{eqn12nllgeq1}). The remaining set of equations 
\begin{equation}
Recur^{12}_{n,0}=\tilde{a}^{12}_{n,0}\;,\quad n\geq0
\end{equation}
give the following identities
\begin{equation}\label{relation12leq0}
\mathfrak{S}^{12}_{n,1}\mathcal{B}^{12}_{n,1}+\mathfrak{U}^{12}_{n-1,1}\mathcal{B}^{12}_{n-1,1}+\mathfrak{W}^{12}_{n,0}\mathcal{B}^{12}_{n,0}=\tilde{a}^{12}_{n,0}\;.
\end{equation}
Because $\mathcal{B}^{12}_{n,\ell}$ have been solved by (\ref{eqn12nllgeq1}) in terms $\mathcal{B}^{12}_{0,\ell}$, the relations (\ref{relation12leq0}) impose infinitely many constraints on $\mathcal{B}^{12}_{0,\ell}$. More precisely, it follows from (\ref{solB12nl}) that $\mathcal{B}^{12}_{n,\ell}$ depends on $\mathcal{B}^{12}_{0,\ell'}$ with $\ell'=\ell,\ell+1,\ldots,\ell+2n$.The $n$-th relation (\ref{relation12leq0}) therefore gives a constraint for the first $2n+2$ coefficients (see Figure \ref{recurconstrs})
\begin{equation}
\mathcal{B}^{12}_{0,\ell}\;,\quad \ell=0,1,\ldots 2n+1\;.
\end{equation}
The number of constraints is ``half as many'' as the number of $\mathcal{B}^{12}_{0,\ell}$.

\subsubsection*{The homogenous equations}
By setting the RHS' of (\ref{Recur12nleqatilde12nl}) to zero, we obtain the homogenous equations 
\begin{equation}\label{Recur12nleq0nl}
\begin{split}
\widetilde{Recur}^{12}_{n,\ell}\equiv\mathfrak{R}^{12}_{n+1,\ell-1}\tilde{\mathcal{B}}^{12}_{n+1,\ell-1}+\mathfrak{S}^{12}_{n,\ell+1}\tilde{\mathcal{B}}^{12}_{n,\ell+1}+\mathfrak{T}^{12}_{n,\ell-1}\tilde{\mathcal{B}}^{12}_{n,\ell-1}{}&\\
+\mathfrak{U}^{12}_{n-1,\ell+1}\tilde{\mathcal{B}}^{12}_{n-1,\ell+1}+\mathfrak{W}^{12}_{n,\ell}\tilde{\mathcal{B}}^{12}_{n,\ell}{}&=0\;,
\end{split}
\end{equation}
\begin{equation}\label{Recur34nleq0nl}
\begin{split}
\widetilde{Recur}^{34}_{n,\ell}\equiv\mathfrak{R}^{34}_{n+1,\ell-1}\tilde{\mathcal{B}}^{34}_{n+1,\ell-1}+\mathfrak{S}^{34}_{n,\ell+1}\tilde{\mathcal{B}}^{34}_{n,\ell+1}+\mathfrak{T}^{34}_{n,\ell-1}\tilde{\mathcal{B}}^{34}_{n,\ell-1}{}&\\
+\mathfrak{U}^{34}_{n-1,\ell+1}\tilde{\mathcal{B}}^{34}_{n-1,\ell+1}+\mathfrak{W}^{34}_{n,\ell}\tilde{\mathcal{B}}^{34}_{n,\ell}{}&=0\;.
\end{split}
\end{equation}
These equations constrain the s-channel decomposition coefficients of a t-channel conformal partial wave, which takes the following form
\begin{equation}
\begin{split}
\frac{(z\bar{z})^{\frac{\Delta_1+\Delta_2}{2}}}{((1-z)(1-\bar{z}))^{\frac{\Delta_2+\Delta_3}{2}}}\Psi^{(t)}_{\Delta_E,\ell_E}(z,\bar{z})={}&\sum_{J=0}^{\infty}\sum_{n=0}^\infty \tilde{\mathcal{B}}^{12}_{n,J} g^{(s)}_{\Delta_1+\Delta_2+2n+J,J}(z,\bar{z})\\
{}&+\sum_{J=0}^{\infty}\sum_{n=0}^\infty \tilde{\mathcal{B}}^{34}_{n,J} g^{(s)}_{\Delta_3+\Delta_4+2n+J,J}(z,\bar{z})\;.
\end{split}
\end{equation}
As in the inhomogeneous case,  the recursion equations 
\begin{equation}
\widetilde{Recur}^{12}_{n,\ell}=\widetilde{Recur}^{34}_{n,\ell}=0\;, \quad \ell\geq1
\end{equation}
allow us to solve $\tilde{\mathcal{B}}^{12}_{n,\ell}$ in terms of $\tilde{\mathcal{B}}^{12}_{0,\ell}$. On the other hand, the remaining equations with $\ell=0$
\begin{equation}
\widetilde{Recur}^{12}_{n,0}=\widetilde{Recur}^{34}_{n,0}=0
\end{equation}
put constraints on the values of $\tilde{\mathcal{B}}^{12}_{0,\ell}$. For example, by looking at the equations with $n=0$ we find that 
simple rational ratios 
\begin{equation}\label{simpleratio12}
\frac{\tilde{\mathcal{B}}^{12}_{0,1}}{\tilde{\mathcal{B}}^{12}_{0,0}}=\frac{1}{(d+\Delta_1+\Delta_2-\Delta_3-\Delta_4)}\left(C^{(2)}_{\Delta_E,\ell_E}+\Delta_1\Delta_2+\frac{\Delta_1\Delta_3(d-\Delta_3)+\Delta_2\Delta_4(d-\Delta_4)}{(\Delta_1+\Delta_2)}\right)\;,
\end{equation} 
\begin{equation}\label{simpleratio34}
\frac{\tilde{\mathcal{B}}^{34}_{0,1}}{\tilde{\mathcal{B}}^{34}_{0,0}}=\frac{1}{(d+\Delta_3+\Delta_4-\Delta_1-\Delta_2)}\left(C^{(2)}_{\Delta_E,\ell_E}+\Delta_3\Delta_4+\frac{\Delta_1\Delta_3(d-\Delta_1)+\Delta_2\Delta_4(d-\Delta_2)}{(\Delta_3+\Delta_4)}\right)
\end{equation} 
for the leading twist coefficients for any spacetime dimension $d>1$.

Before we end this subsection let us consider a special case where the t-channel conformal partial wave has $\ell_E=0$, and the external weights are degenerate $\Delta_i=\Delta_\phi$. In this case the solution to (\ref{Recur12nleq0nl}) and (\ref{Recur34nleq0nl}) has the interpretation of anomalous dimensions (see Appendix \ref{appeqweight}). The seed coefficients for the scalar exchange was obtained first in \cite{Giombi:2018vtc}\footnote{A straightforward exercise is to check that the expressions satisfy the relations (\ref{simpleratio12}), (\ref{simpleratio34}).}
\begin{equation}
\begin{split}
\tilde{\mathcal{B}}^{12}_{0,\ell}={}&\frac{4 \Gamma (\Delta_E) \Gamma \left(\frac{d-\Delta_E}{2}\right)^2 \Gamma (d+\ell -2) \Gamma (\ell +\Delta_\phi )^2 \Gamma (\ell +2 \Delta_\phi -1)}{\Gamma (d-1) \Gamma \left(\frac{\Delta_E}{2}\right)^2 \Gamma (\Delta_\phi )^2 \Gamma (\ell +1) \Gamma \left(\frac{d}{2}-\Delta_E\right) \Gamma \left(\frac{d}{2}+\ell -1\right) \Gamma (2 \ell +2 \Delta_\phi -1)}\\
{}&\times p_\ell\left[i\left(\frac{\Delta_E}{2}-\frac{d}{4}\right);\frac{d}{4},\Delta_\phi-\frac{d}{4},\Delta_\phi-\frac{d}{4},\frac{d}{4}\right]
\end{split}
\end{equation}
where $p_\ell(x;\mathfrak{a},\mathfrak{b},\mathfrak{c},\mathfrak{d})$ is the Wilson polynomial defined in (\ref{Wilsonpolyn}),\footnote{These Wilson polynomials also appear in the crossed channel decomposition of conformal partial waves of spinning operators \cite{Sleight:2018ryu}.} and we will suppress its argument in the following to write it as $p_\ell$. With this input, our method allows us to efficiently compute all the subleading coefficients. From $\widetilde{Recur}^{12}_{0,\ell}=0$, we find 
\begin{equation}\small
\begin{split}
{}&\tilde{\mathcal{B}}^{12}_{1,\ell}=-\frac{2^{\Delta_E-2 \Delta_\phi -2 \ell -2}\Gamma \left(\frac{\Delta_E+1}{2}\right) \Gamma \left(\frac{d-\Delta_E}{2}\right)^2  \Gamma (\ell +\Delta_\phi +1) \Gamma (\ell +2 \Delta_\phi )}{\Gamma \left(\frac{d}{2}\right) \Gamma \left(\frac{\Delta_E}{2}\right) \Gamma (\Delta_\phi )^2 \Gamma (\ell +1) \Gamma \left(\frac{d}{2}-\Delta_E\right) \Gamma \left(\ell +\Delta_\phi +\frac{3}{2}\right)}\\{}&\times\bigg[\frac{(\Delta_\phi +\ell ) (-d+2 \Delta_\phi +\ell +2) (-d+4 \Delta_\phi +2 \ell )}{-d+4 \Delta_\phi +2 \ell +2}p_{\ell}+\frac{(d+\ell -1) (d+2 \ell +2) (\Delta_\phi +\ell +1) (2 \Delta_\phi +\ell )}{(\ell +1) (d+2 \ell )}p_{\ell+2}\\
{}&-\frac{(2 \Delta_\phi +2 \ell +1) (C^{(2)}_{\Delta_E,0}+\Delta_\phi  (d+2 \ell +2)+\ell ^2+\ell)}{\ell +1}p_{\ell+1}\bigg]\;.
\end{split}
\end{equation}
Substituting the solution into $\widetilde{Recur}^{12}_{1,\ell}=0$, we get an expression for $\tilde{\mathcal{B}}^{12}_{2,\ell}$ which has the following form 
\begin{equation}
\tilde{\mathcal{B}}^{12}_{2,\ell}=(\ldots)p_{\ell}+(\ldots)p_{\ell+1}+(\ldots)p_{\ell+2}+(\ldots)p_{\ell+3}+(\ldots)p_{\ell+4}\;.
\end{equation}
The coefficients in this expression are more complicated and we will refrain from writing down their explicit expressions. The above two sets of subleading coefficients were computed in \cite{Sleight:2018ryu} using a different method. We have checked that these expressions are equivalent to their results.\footnote{We thank Massimo Taronna for providing the Mathematica notebook that contains their relevant results.} It is totally straightforward to iterate and get $\tilde{\mathcal{B}}^{12}_{n,\ell}$ with higher $n$. The result is expressed as a linear combination of $p_{\ell}$, $p_{\ell+1}$, \ldots $p_{\ell+2n+2}$, and the coefficients can be efficiently computed using the recursion relations. However, in contrast to the 1d case, we have not found an obvious way to write the coefficients $\tilde{\mathcal{B}}^{12}_{n,\ell}$ for any $n$ in a closed form.

\section{Discussion and Outlook}\label{SecDiscuss}
In this paper we performed a systematic position space analysis of the conformal block decomposition of Witten diagrams and conformal partial waves. In our analysis we emphasized the use of the equation of motion operator and the contact Witten diagrams. These objects played important roles in the decomposition of exchange Witten diagrams, both in the direct channel and in the crossed channel. Our main finding is a recursive algorithm for computing the crossed channel decomposition coefficients of exchange Witten diagrams and conformal partial waves. This algorithm allows us to efficiently obtain the OPE coefficient of any double-trace operator with sub-leading conformal twist, in terms of the coefficients of the double-trace operators with the minimal twist. At face value, our results provide a useful tool to extract CFT data from tree-level holographic correlators for all internal spins, especially when the exchange Witten diagrams do not admit a truncated representation in terms of finitely many $D$-functions.

Let us also mention other problems where our results might be useful.
\subsubsection*{Mellin bootstrap}
One use of our results is to simplify the Mellin bootstrap program. The Mellin bootstrap approach uses  the crossing symmetrized exchange Witten diagrams as an expansion basis. The use of the exchange Witten diagrams introduces spurious double-trace operators in the conformal block decomposition of the correlator ansatz. The method hence derives the bootstrap conditions by requiring the vanishing of all the double-trace coefficients when summing over the physical spectrum. In the Mellin bootstrap method one obtains the double-trace coefficients of a single Witten diagram by taking residues of the Mellin amplitude at the double-trace poles, and then projecting them to different spins using continuous Hahn polynomials. As we already mentioned in the introduction, the projection at a certain pole gives only the mixed OPE coefficients between the primary double-trace operators and the descendant double-trace operators for which the primaries have smaller twists. For the Mellin bootstrap method, it is not necessary to solve the mixing problem. This is because the descendant contributions always vanish in the bootstrap equations if the conditions on the primary operators are already satisfied. Nevertheless, it might be useful to clarify the structures of the Mellin bootstrap equations by eliminating the redundant descendant contributions. Using our method, such contributions are absent automatically because we work with the primary operators only. Furthermore, because the decomposition coefficients satisfy recursion relations, it should be possible to recursively derive sub-leading bootstrap conditions from the leading ones. These sub-leading conditions are crucial for probing operators with sub-leading twists, for example in the $4-\epsilon$ expansion \cite{Gopakumar:2018xqi}. One should however notice that there still is a subtlety in the Mellin bootstrap method which is to fix the contact term ambiguity in the basis (see \cite{Dey:2017fab,Gopakumar:2018xqi} for recent progress in general $d$, and \cite{Mazac:2018ycv} for $d=1$). This issue must be  addressed separately.

\subsubsection*{Analytic functionals in $\mathrm{CFT}_1$}
It was pointed out in \cite{Mazac:2018ycv} that the decomposition coefficients of the following crossing symmetric combination of $AdS_2$ Witten diagrams\footnote{In \cite{Mazac:2018ycv} the external operators are restricted to be identical and all have the same conformal dimension $\Delta_\phi$. The coefficient $\lambda$ can be fixed such that the coefficient of $\partial g^{(s)}_{2\Delta_\phi}$ is zero in the s-channel decomposition. This is one of the infinitely many equivalent choices of $\lambda$, see \cite{Mazac:2018ycv} for more details.}
\begin{equation}
W^{s,exchange}_\Delta+W^{t,exchange}_\Delta+W^{u,exchange}_\Delta+\lambda W^{contact}
\end{equation}
encode the information of the complete set of functionals for $\mathrm{CFT}_1$ (see \cite{Mazac:2016qev,Mazac:2018mdx} for earlier related work, also \cite{Mazac1dinversion}). A basis of analytic functionals is given by $\alpha_n$ and $\beta_m$, labelled by integers $n=0,1,2,\ldots$, $m=1,2,3,\ldots$, and the application of the functionals to the crossing equation gives the complete set of constraints. It is sufficient to know the action of the functionals on the bootstrap vector function $F_\Delta(z)$ defined by
\begin{equation}
F_\Delta(z)=\frac{g^{(s)}_\Delta(z)}{z^{2\Delta_\phi}}-\frac{g^{(s)}_\Delta(1-z)}{(1-z)^{2\Delta_\phi}}\;.
\end{equation}
The action of the $n$-th functional on the function $F_\Delta$ can be expressed as the ratio of decomposition coefficients of the Witten diagrams
\begin{equation}\label{functionalaction}
\begin{split}
\alpha_n[F_\Delta]={}&-\frac{A_n+2B_{2n}+\lambda a_n}{A}\;,\\
\beta_n[F_\Delta]={}&-\frac{C^{(s)}_n+2C^{(t)}_{2n}+\lambda c_n}{A}\;.
\end{split}
\end{equation}
We have computed all these decomposition coefficients in this paper, and we briefly remind the reader of our notations. $A$, $A_n$, $C^{s}_n$ are respectively the coefficient of $g^{(s)}_\Delta$, $g^{(s)}_{2\Delta_\phi+2n}$ and $\partial g^{(s)}_{2\Delta_\phi+2n}$ of  the exchange diagram $W^{s,exchange}_\Delta$ in the s-channel\footnote{The explicit expressions are given in (\ref{DirectcoeA}), (\ref{DirectcoeAn}), (\ref{DirectcoeCns}).}; $B_{2n}$, $C^{(t)}_{2n}$ are the coefficient of $g^{(s)}_{2\Delta_\phi+2n}$ and $\partial g^{(s)}_{2\Delta_\phi+2n}$ of the exchange diagrams $W^{t,exchange}_\Delta$  or $W^{u,exchange}_\Delta$ in the t and u-channel; and finally $a_n$ and $c_n$ are the coefficient of $g^{(s)}_{2\Delta_\phi+2n}$ and $\partial g^{(s)}_{2\Delta_\phi+2n}$ of the contact diagram $W^{contact}=D_{\Delta_\phi\Delta_\phi\Delta_\phi\Delta_\phi}$. In \cite{Mazac:2018ycv}, the actions of the functionals are constructed as contour integrals against certain weight functions. For general $\Delta_\phi$ and $n$, evaluating these integrals to yield explicit expressions is still technically challenging. On the other hand, our methods for computing the decomposition coefficients of Witten diagrams have no restrictions on quantum numbers.
 By exploiting the relation (\ref{functionalaction}) between the two, our techniques therefore provide a complementary way to compute the analytic functional actions. Our results for the decomposition coefficients can be easily assembled to give the general analytic functionals for arbitrary external dimension $\Delta_\phi$, and recursively for all $n$, which should be particularly useful for the numeric bootstrap application using the analytic functionals \cite{Paulos1dnumeric}.

\subsubsection*{$6j$ symbols}
Moreover, the recursive algorithm we formulated here may provide some help to the computation of the $6j$ symbol in general dimensions. In \cite{Liu:2018jhs}, the $6j$ symbols in $d=1,2,4$ were computed using the Lorentzian inversion formula \cite{Caron-Huot:2017vep}, and expressed in terms of hypergeometric functions ${}_4F_3$. However evaluating $6j$ symbols in other dimensions still remains a challenging open problem.\footnote{The $6j$ symbols  also admit an Mellin Barnes integral representation \cite{Sleight:2018ryu}. The integral representation is in general quite complicated but simplifies in certain cases.} This is due to the fact that the explicit form of the conformal partial waves is not known in odd spacetime dimensions, while in even dimensions $d>4$ the Lorentzian inversion integral does not factorize. The $6j$ symbols are intimately related to the crossed channel decomposition of conformal partial waves. More precisely the $6j$ symbol can be viewed as the spectral density function for decomposing an t-channel conformal partial wave into the s-channel conformal partial waves. By eliminating the s-channel shadow conformal block and closing the contour, the encircled poles of the $6j$ symbol correspond to the double-trace operators and their residues give the crossed channel decomposition coefficients. Since we can in principle obtain all the double-trace OPE coefficients using our recursive algorithm, once we input the coefficients of the leading twist double-trace operators ({\it e.g.} from Mellin space), it is natural to ask the following question: knowing the poles and residues, is there a convenient way to reverse engineer the $6j$ symbol? We will not further explore this problem in this paper, but the apparent advantage of such a method is that it is insensitive to the spacetime dimensions. 

\subsubsection*{Boundary conformal field theories}
The techniques we discussed in this paper also admit a straightforward extension to boundary conformal field theories. The simplest holographic setup for BCFT is obtained by taking a half of the $AdS_{d+1}$ space which ends on a $AdS_d$ boundary. We further require fields in $AdS_{d+1}$ to satisfy Neumann boundary condition on $AdS_d$. Two-point functions on the conformal boundary $AdS_{d+1}$ now become the simplest non-trivial objects to study. There are two types of exchange Witten diagrams for two-point functions, namely, the bulk channel exchange Witten diagram and the boundary channel exchange Witten diagrams. These exchange Witten diagrams have similar decomposition properties to those of the four-point functions, {\it i.e.}, only double-trace operators appear in the crossed channel and both the single-trace operator and double-trace operators appear in the direct channel \cite{Rastelli:2017ecj}. The equation of motion operators and properties of conformal blocks allow us to similarly formulate recursive algorithms for solving the crossed channel decomposition coefficients of exchange Witten diagrams. We discuss the BCFT extension in a separate publication \cite{bcftpolyakov}, where we also use the decomposition coefficients to perform Polyakov style bootstrap \cite{Polyakov:1974gs}.

There are also other extensions worth exploring.  One is to repeat the analysis for four-point Witten diagrams with external spinning operators. Another question is whether one can also find similar recursive techniques for decomposing AdS loop diagrams. Finally, it would also be interesting to incorporate supersymmetry which presumably will further facilitate the extraction of CFT data from holographic correlators in supersymmetric backgrounds.    

\acknowledgments
I thank Dalimil Maz\'a\v{c} and for discussions and collaboration on a related project \cite{bcftpolyakov}. I am grateful to Dalimil Maz\'a\v{c}, Wolfger Peelaers, Leonardo Rastelli and especially Massimo Taronna for carefully reading the draft and helpful comments.  I also thank  Rajesh Gopakumar, Eric Perlmutter, Jo\~ao Penedones, David Simmons-Duffin, Massimo Taronna, Balt van Rees and other participants of the Bootstrap 2018 for useful conversations and comments on the work. I wish to thank the California Institute of Technology for the great hospitality during the Bootstrap 2018 workshop where part of this work was finished. This work is supported in part by NSF Grant PHY-1620628.

\appendix
\section{Contact Diagrams}\label{appcontact}
As was explained in \cite{Penedones:2010ue}, a generic contact Witten diagram can be written as linear combinations of (\ref{Dnijdef}), and the calculation can be streamlined using the embedding space formalism. We now want to prove that among all the $D^{\{n_{ij}\}}_{\Delta_1\Delta_2\Delta_3\Delta_4}$, the ones with $n_{13}=n_{24}=n_{23}=n_{34}=0$ form a basis. It is convenient to show this using the Mellin representation formalism \cite{Mack:2009mi,Penedones:2010ue}. 
In this formalism, a scalar correlator\footnote{Strictly speaking we should take the connected part of the correlator.} with external dimensions $\Delta_i$ is represented as a multi-dimensional inverse Mellin integral 
\begin{equation}\label{GMellinrep}
G(x_i)=\int [d\delta_{ij}] \left(\prod_{i<j}(x_{ij}^2)^{-\delta_{ij}}\Gamma(\delta_{ij})\right)\mathcal{M}(\delta_{ij})
\end{equation}
where $\mathcal{M}(\delta_{ij})$ is called the Mellin amplitude. The $\delta_{ij}$ are symmetric $\delta_{ij}=\delta_{ji}$, and satisfy
\begin{equation}
\sum_{j\neq i}\delta_{ij}=\Delta_i\;,
\end{equation}
as a result of conformal covariance. In the case of four-point functions, only two $\delta_{ij}$ are independent. It is convenient to use the following parameterization in terms of $s$ and $t$
\begin{equation}
\begin{split}
{}&\delta_{12}=\frac{\Delta_1+\Delta_2-s}{2}\;,\quad \delta_{34}=\frac{\Delta_3+\Delta_4-s}{2}\;,\\
{}&\delta_{13}=\frac{\Delta_1+\Delta_3-t}{2}\;,\quad \delta_{24}=\frac{\Delta_2+\Delta_4-t}{2}\;,\\
{}&\delta_{14}=\frac{s+t-\Delta_2-\Delta_3}{2}\;,\quad \delta_{23}=\frac{s+t-\Delta_1-\Delta_4}{2}\;.
\end{split}
\end{equation}
In the Mellin formalism, $D$-functions in (\ref{Dfunction}) admit particularly simple representation -- their Mellin amplitudes are simply constants \cite{Penedones:2010ue}
\begin{equation}
\mathcal{M}_{D_{\Delta_1\Delta_2\Delta_3\Delta_4}}(s,t)=\frac{\pi^{\frac{d}{2}}\Gamma[\frac{\sum_{i}\Delta_i-d}{2}]}{\prod_i \Gamma[\Delta_i]}\;.
\end{equation} 
Using the definition (\ref{Dnijdef}) and the above Mellin representation for $D$-functions, it is easy to obtain the Mellin representation for $D^{\{n_{ij}\}}_{\Delta_1\Delta_2\Delta_3\Delta_4}$ 
\begin{equation}
\begin{split}
D^{\{n_{ij}\}}_{\Delta_1\Delta_2\Delta_3\Delta_4}={}&\prod_{i<j}(x_{ij}^2)^{n_{ij}}D_{\Delta_1^{n_{ij}}\Delta_2^{n_{ij}}\Delta_3^{n_{ij}}\Delta_4^{n_{ij}}}\\
={}&\int [d\delta_{ij}] \left(\prod_{i<j}(x_{ij}^2)^{-\delta_{ij}}\Gamma(\delta_{ij})\right)\underbrace{\left(\prod_{i<j}\left(\delta_{ij}\right)_{n_{ij}}\mathcal{M}_{D_{\Delta_1^{n_{ij}}\Delta_2^{n_{ij}}\Delta_3^{n_{ij}}\Delta_4^{n_{ij}}}}
\right)}_{\mathcal{M}_{D^{\{n_{ij}\}}_{\Delta_1\Delta_2\Delta_3\Delta_4}}(s,t)}\;.
\end{split}
\end{equation}
Clearly the Mellin amplitude $\mathcal{M}_{D^{\{n_{ij}\}}_{\Delta_1\Delta_2\Delta_3\Delta_4}}(s,t)$ of $D^{\{n_{ij}\}}_{\Delta_1\Delta_2\Delta_3\Delta_4}$ is a polynomial of $s$ and $t$ of degree $\sum_{i<j}n_{ij}$ \cite{Penedones:2010ue}. Since a contact Witten diagram can always be expressed as a linear combination of $D^{\{n_{ij}\}}_{\Delta_1\Delta_2\Delta_3\Delta_4}$, the Mellin amplitude of any contact Witten diagram is also a polynomial.

Obviously, any polynomial in $s$ and $t$ can be formed from the linear combination of the monomials
\begin{equation}
(\delta_{12})_{n_{12}}=\left(\frac{\Delta_1+\Delta_2-s}{2}\right)_{n_{12}}\;,\quad (\delta_{14})_{n_{14}}=\left(\frac{s+t-\Delta_2-\Delta_3}{2}\right)_{n_{14}}\;,
\end{equation}
where the integers in the $n_{12}$, $n{14}$  are restricted to be $n_{12}, n_{14}\geq 0$. This implies that the Mellin amplitudes of $D^{\{n_{ij}\}}_{\Delta_1\Delta_2\Delta_3\Delta_4}$ with $n_{13}=n_{24}=n_{23}=n_{34}=0$ form a basis for the Mellin amplitude of any contact diagram. This statement is equivalent to the position space statement that these $D^{\{n_{ij}\}}_{\Delta_1\Delta_2\Delta_3\Delta_4}$ form a basis for any four-point contact Witten diagram.

Now let us now return to address the question raised in footnote \ref{footnoteEOMonContact}, and prove that acting on a contact Witten diagram with the equation of motion produces finitely many contact Witten diagrams. To achieve this, we write the operator $\mathbf{EOM}^{(s)}$ in (\ref{EOMszzb}) as an differential operator of the cross ratios $U$ and $V$
\begin{equation}
\begin{split}
\mathbf{EOM}^{(s)}={}&2(UV^{-1}+1-V^{-1})V\frac{\partial}{\partial V}\left(V\frac{\partial}{\partial V}+a+b\right)-2U\frac{\partial}{\partial U}\left(2U\frac{\partial}{\partial U}-d\right)\\
{}&+2(1+U-V)\left(U\frac{\partial}{\partial U}+V\frac{\partial}{\partial V}+a\right)\left(U\frac{\partial}{\partial U}+V\frac{\partial}{\partial V}+b\right)+C^{(2)}_{\Delta_E,\ell_E}\;.
\end{split}
\end{equation}
On the other hand, using (\ref{GMellinrep}) we have the following Mellin representation for the contact Witten diagram $\mathcal{W}^{contact}(U,V)$
\begin{equation}
\begin{split}
\mathcal{W}^{contact}(U,V)={}&\int_{-i\infty}^{i\infty}\frac{ds}{2}\frac{dt}{2}U^{\frac{s}{2}}V^{\frac{t}{2}-\frac{\Delta_2+\Delta_3}{2}}\mathcal{M}^{contact}(s,t)\Gamma[\frac{\Delta_1+\Delta_2-s}{2}]\Gamma[\frac{\Delta_3+\Delta_4-s}{2}]\\
{}&\times \Gamma[\frac{\Delta_1+\Delta_4-t}{2}]\Gamma[\frac{\Delta_2+\Delta_3-t}{2}]\Gamma[\frac{s+t-\Delta_1-\Delta_3}{2}]\Gamma[\frac{s+t-\Delta_2-\Delta_4}{2}]
\end{split}
\end{equation}
where the Mellin amplitude $\mathcal{M}^{contact}(s,t)$ is a polynomial. It is easy to see the differential operators in $\mathbf{EOM}^{(s)}$ can be interpreted as difference operators acting on the Mellin amplitude, according to the following dictionary
\begin{equation}
\begin{split}
U\frac{\partial}{\partial U}\to{}& \frac{s}{2}\times\;,\\
V\frac{\partial}{\partial V}\to{}& \left(\frac{t}{2}-\frac{\Delta_2+\Delta_3}{2}\right)\times\;,\\
U^mV^n\to{}& \mathcal{M}(s-2m,t-2n)\left(\frac{\Delta_1+\Delta_2-s}{2}\right)_m\left(\frac{\Delta_3+\Delta_4-s}{2}\right)_m\left(\frac{\Delta_1+\Delta_4-t}{2}\right)_n\\
{}&\times \left(\frac{\Delta_2+\Delta_3-t}{2}\right)_n\left(\frac{s+t-\Delta_1-\Delta_3}{2}\right)_{-m-n}\left(\frac{s+t-\Delta_2-\Delta_4}{2}\right)_{-m-n}\;.
\end{split}
\end{equation}
Acting on $\mathcal{M}^{contact}(s,t)$ with $\mathbf{EOM}^{(s)}$ interpreted as a difference operator, we find that the result is still a polynomial. The factors $(UV^{-1}+1-V^{-1})$, $(1+U-V)$ could have introduced poles in $s$ and $t$ from the Pochhammer symbols, but these poles are precisely cancelled by the $s$, $t$ polynomials introduced by $V\frac{\partial}{\partial V}\left(V\frac{\partial}{\partial V}+a+b\right)$ and $\left(U\frac{\partial}{\partial U}+V\frac{\partial}{\partial V}+a\right)\left(U\frac{\partial}{\partial U}+V\frac{\partial}{\partial V}+b\right)$. This concludes the proof that a contact Witten diagram  acted on by the equation of motion operator can again be written as a finite linear combination of contact Witten diagrams.

\section{The Special Case of Equal Weights}\label{appeqweight}
As we explained in footnote \ref{FNderivativeblock}, derivative conformal blocks appear in the s-channel decomposition when $\Delta_1+\Delta_2=\Delta_3+\Delta_4+2m$, $m\in \mathbb{Z}$ as a consequence of large $N$ counting. We can also see this explicitly from the expressions of the decomposition coefficients. Let us take the zero-derivative contact diagram (\ref{Dfunctioningxi}) as a concrete example. Without loss of generality, we assume that $\Delta_1+\Delta_2\geq \Delta_3+\Delta_4$. Both coefficients $a^{12}_{n,0}$ and $a^{34}_{n',0}$ contain simple poles when $\Delta_1+\Delta_2-\Delta_3-\Delta_4$ is an even integer, which come from the Gamma factors $\Gamma(\frac{-2n-\Delta_1-\Delta_2+\Delta_3+\Delta_4}{2})$ and $\Gamma(\frac{-2n'-\Delta_3-\Delta_4+\Delta_1+\Delta_2}{2})$ respectively. More precisely, let $\Delta_1+\Delta_2-\Delta_3-\Delta_4=2m+\eta$ where $m$ is a non negative integer and $\eta$ is small, then $a^{12}_{n,0}$ has a simple pole $\propto 1/\eta$ for any non negative integer $n$ while $a^{34}_{n',0}$ has a simple pole $\propto 1/\eta$ when $n'\geq m$. Moreover we can check that the residue of the $1/\eta$ pole is the same for $a^{12}_{n,0}$ and $a^{34}_{n+m,0}$ up to a flipped sign. The limit of $\eta\to 0$ gives rise to the derivative blocks
\begin{equation}
\begin{split}
{}&a^{12}_{n,0}g^{(s)}_{\Delta_1+\Delta_2+2n+J,J}+a^{34}_{n,0}g^{(s)}_{\Delta_3+\Delta_4+2n+2m+J,J}\\
{}&\propto \left(\frac{1}{\eta}+c_1+\mathcal{O}(\eta)\right)g^{(s)}_{\Delta_1+\Delta_2+2n+J,J}+\left(-\frac{1}{\eta}+c_2+\mathcal{O}(\eta)\right)g^{(s)}_{\Delta_1+\Delta_2+2n-\eta+J,J}\\
{}&\xrightarrow{\eta\to 0} \partial g^{(s)}_{\Delta_1+\Delta_2+2n+J,J}+(c_1+c_2)g^{(s)}_{\Delta_1+\Delta_2+2n+J,J}\;.
\end{split}
\end{equation}
Therefore when $\Delta_1+\Delta_2-\Delta_3-\Delta_4=2m\geq0$, the expansion of the contact diagram becomes
\begin{equation}\label{WcontactinsEW}
W^{contact}(x_i)=\sum_{J=0}^{J_{\rm max}}\sum_{n=0}^\infty a_{n,J} g^{(s)}_{\Delta_1+\Delta_2-2m+2n+J,J}(x_i)+\sum_{J=0}^{J_{\rm max}}\sum_{n=0}^\infty c_{n,J} \partial g^{(s)}_{\Delta_1+\Delta_2+2n+J,J}(x_i)\;
\end{equation}
where the coefficients $a_{n,J}$ and $b_{n,J}$ are obtained from $a^{12}_{n,J}$ and $a^{34}_{n,J}$ by taking the limit as we did above. Moreover, the coefficients of the derivative blocks can be interpreted as the anomalous dimensions of the double-trace operators. This is clear from expanding the conformal block decomposition of a correlator to the first order in anomalous dimensions
\begin{equation}
\begin{split}
G(x_i)={}&\ldots+\sum_{J=0}^{J_{\rm max}}\sum_{n=0}^\infty a_{n,J}^{h} g^{(s)}_{\Delta_1+\Delta_2+2n+\gamma_{n,J} h+J,J}(x_i)\\
={}&\ldots+\sum_{J=0}^{J_{\rm max}}\sum_{n=0}^\infty a_{n,J}^{(0)}g^{(s)}_{\Delta_1+\Delta_2+2n+J,J}(x_i)\\
{}&+h\left(a_{n,J}^{(1)}g^{(s)}_{\Delta_1+\Delta_2+2n+J,J}(x_i)+\underbrace{a_{n,J}^{(0)}\gamma_{n,J}}_{c_{n,J}} \partial g^{(s)}_{\Delta_1+\Delta_2+2n+J,J}(x_i)\right)+\mathcal{O}(h^2)
\end{split}
\end{equation}
where $a^{(i)}_{n,J}=\frac{1}{i!}\frac{\partial^i}{\partial h^i}a^{h}_{n,J}$ and $h$ is just a placeholder for the small parameter.

The exchange Witten diagrams admit similar s-channel expansions   when the external conformal dimensions are fine-tuned to satisfy $\Delta_1+\Delta_2-\Delta_3-\Delta_4\in2\mathbb{Z}$. For simplicity, we will focus on the equal weight case, {\it i.e.}, $\Delta_i=\Delta_\phi$. Then the exchange diagrams have the following decompositions
\begin{equation}\label{WexchangeinsEW}
W^{s,\,exchange}_{\Delta_E,\ell_E}=A\, g^{(s)}_{\Delta_E,\ell_E}(x_i) +\sum_{J=0}^{\ell_E}\sum_{n=0}^\infty A_{n,J} g^{(s)}_{2\Delta_\phi+2n+J,J}(x_i)+\sum_{J=0}^{\ell_E}\sum_{n=0}^\infty C_{n,J}^{(s)} \partial g^{(s)}_{2\Delta_\phi+2n+J,J}(x_i)\;,
\end{equation}
in the direct channel, and
\begin{equation}\label{WexchangeintEW}
W^{t,\,exchange}_{\Delta_E,\ell_E}=\sum_{J=0}^{\ell_E}\sum_{n=0}^\infty B_{n,J} g^{(s)}_{2\Delta_\phi+2n+J,J}(x_i)+\sum_{J=0}^{\ell_E}\sum_{n=0}^\infty C_{n,J}^{(t)} \partial g^{(s)}_{2\Delta_\phi+2n+J,J}(x_i)\;,
\end{equation}
in the crossed channel. In principle, we can obtain these decomposition coefficients from the generic case by taking the equal weight limit. This is not difficult to perform on the direct channel decomposition coefficients since the coefficients with generic $\Delta_i$ can be written down in a closed form for all $n$ and $J$. For example, when the exchanged field is a scalar, {\it i.e.}, $\ell_E=0$, application of the strategy in Section \ref{Secdirectchan} gives 
\begin{equation}\small
\begin{split}
A={}&\frac{\pi ^{d/2} \Gamma \left(\frac{\Delta_E +\Delta_1-\Delta_2}{2}\right) \Gamma \left(\frac{\Delta_E -\Delta_1+\Delta_2}{2}\right) \Gamma \left(\frac{-\Delta_E +\Delta_1+\Delta_2}{2}\right) \Gamma \left(\frac{\Delta_E +\Delta_3-\Delta_4}{2}\right) \Gamma \left(\frac{\Delta_E -\Delta_3+\Delta_4}{2}\right) \Gamma \left(\frac{-\Delta_E +\Delta_3+\Delta_4}{2}\right)}{ \Gamma \left(-\frac{d}{2}+\Delta_E +1\right)}\\
{}&\times \frac{\Gamma \left(\frac{-d+\Delta_E +\Delta_1+\Delta_2}{2}\right) \Gamma \left(\frac{-d+\Delta_E +\Delta_3+\Delta_4}{2}\right)}{8 \Gamma (\Delta_E ) \Gamma (\Delta_1) \Gamma (\Delta_2) \Gamma (\Delta_3) \Gamma (\Delta_4)}\;,
\end{split}
\end{equation}
and $A_{n,J}^{12}=A^{12}_n\delta_{0,J}$, with
\begin{equation}\small
\begin{split}
A^{12}_n={}&\frac{2 \pi ^{d/2} (-1)^n \Gamma (n+\Delta_1) \Gamma (n+\Delta_2) \Gamma \left(-\frac{d}{2}+n+\Delta_1+\Delta_2\right) \Gamma \left(\frac{2 n+\Delta_1+\Delta_2+\Delta_3-\Delta_4}{2}\right) }{n! \Gamma (\Delta_1) \Gamma (\Delta_2) \Gamma (\Delta_3) \Gamma (\Delta_4) \Gamma (2 n+\Delta_1+\Delta_2) }\\
{}&\times\frac{\Gamma \left(\frac{2 n+\Delta_1+\Delta_2-\Delta_3+\Delta_4}{2}\right) \Gamma \left(\frac{-2 n-\Delta_1-\Delta_2+\Delta_3+\Delta_4}{2}\right) \Gamma \left(\frac{-d+2 n+\Delta_1+\Delta_2+\Delta_3+\Delta_4}{2}\right)}{\left((d-2 \Delta_E )^2-4 \left(-\frac{d}{2}+\Delta_1+\Delta_2+2 n\right)^2\right) \Gamma \left(-\frac{d}{2}+2 n+\Delta_1+\Delta_2\right)}\;.
\end{split}
\end{equation}
The coefficients $A^{34}_n$ are related to $A^{12}_n$ by exchanging $\Delta_1$, $\Delta_2$ with $\Delta_3$, $\Delta_4$. By further taking the equal weight limit, we obtain the coefficients $A$, $A_{n,J}=A_n\delta_{0,J}$, $C_{n,J}^{(s)}=C_n^{(s)}\delta_{0,J}$. The explicit expressions are recorded here for reader's convenience 
\begin{equation}\label{DirectcoeA}\small
A=\frac{\pi ^{d/2} \Gamma \left(\frac{\Delta_E}{2}\right)^4 \Gamma \left(\Delta_\phi -\frac{\Delta_E}{2}\right)^2 \Gamma \left(\frac{\Delta_E-d}{2}+\Delta_\phi \right)^2}{8 \Gamma (\Delta_E) \Gamma (\Delta_\phi )^4 \Gamma \left(-\frac{d}{2}+\Delta_E+1\right)}
\end{equation}
\begin{equation}\small\label{DirectcoeAn}
\begin{split}
A_n={}&\frac{\pi ^{d/2} \Gamma (n+\Delta_\phi )^4 \Gamma \left(-\frac{d}{2}+n+2 \Delta_\phi \right)^2 }{(n!)^2 \Gamma (\Delta_\phi )^4 (\Delta_E -2 \Delta_\phi -2 n)^2 \Gamma (2 (n+\Delta_\phi )) (-d+\Delta_E +2 \Delta_\phi +2 n)^2 \Gamma \left(2 (n+\Delta_\phi )-\frac{d}{2}\right)}\\
{}&\times\bigg(d-4 (\Delta_\phi +n)+(-\Delta_E +2 \Delta_\phi +2 n) (d-\Delta_E -2 \Delta_\phi -2 n)\bigg[-\psi\left(-\frac{d}{2}+n+2 \Delta_\phi \right)\\
{}&+\psi\left(2 (n+\Delta_\phi )-\frac{d}{2}\right)-2 \psi(n+\Delta_\phi )+\psi(2 (n+\Delta_\phi ))+\psi(n+1)\bigg]\bigg)
\end{split}
\end{equation}
\begin{equation}\small\label{DirectcoeCns}
C_n^{(s)}=\frac{\pi ^{d/2} \Gamma (n+\Delta_\phi )^4 \Gamma \left(-\frac{d}{2}+n+2 \Delta_\phi \right)^2}{(n!)^2 \Gamma (\Delta_\phi )^4 (-\Delta_E +2 \Delta_\phi +2 n) \Gamma (2 (n+\Delta_\phi )) (-d+\Delta_E +2 \Delta_\phi +2 n) \Gamma \left(2 (n+\Delta_\phi )-\frac{d}{2}\right)}
\end{equation}
Here $\psi(z)=\Gamma'(z)/\Gamma(z)$ is the digamma function.

For the crossed channel decomposition, however, it is more convenient to obtain recursion relations for $B_{n,J}$ and $C^{(t)}_{n,J}$ whose solution is determined by the $n=0$ data. In this way, we only need to take the equal weight limit once for $n=0$. In the rest of this appendix, we will give the explicit expressions for these recursion relations, first for $d>1$ and then for $d=1$.

\subsubsection*{Equal weight recursion relations for $d>1$}
We first need the action of the operator $\mathbf{EOM}^{(t)}$ on the derivative block $\partial_\Delta F_{\lambda_1,\lambda_2}$. This can be obtained from (\ref{EOMrecuranyd}) by taking derivative with respect to $\Delta$.\footnote{Note that $\Delta$ does not appear in the differential operator $\mathbf{EOM}^{(t)}$.} The result is
\begin{equation}
\begin{split}
\mathbf{EOM}^{(t)}[\partial_\Delta\mathcal{F}_{\lambda_1,\lambda_2}]={}&\mathcal{R}\, F_{\lambda_1,\lambda_2-1}+\mathcal{S}\,F_{\lambda_1-1,\lambda_2}+\mathcal{T}\,F_{\lambda_1+1,\lambda_2}+\mathcal{U}\,F_{\lambda_1,\lambda_2+1}+\mathcal{W}\,F_{\lambda_1,\lambda_2}\\
{}&+\mathcal{R}'\, F_{\lambda_1,\lambda_2-1}+\mathcal{S}'\,F_{\lambda_1-1,\lambda_2}+\mathcal{T}'\,F_{\lambda_1+1,\lambda_2}+\mathcal{U}'\,F_{\lambda_1,\lambda_2+1}+\mathcal{W}'\,F_{\lambda_1,\lambda_2}\;.
\end{split}
\end{equation}
where
\begin{equation}
\begin{split}
{}&\mathcal{R}=-\frac{(\ell +2 \epsilon ) (-\Delta +2 \Delta_\phi +\ell )^2}{2 (\ell +\epsilon )}\;,\\
{}&\mathcal{S}=-\frac{\ell  (\Delta -2 \Delta_\phi +\ell +2 \epsilon )^2}{2 (\ell +\epsilon )}\;,\\
{}&\mathcal{T}=-\frac{(\Delta -1) (\Delta -2 \epsilon ) (\Delta +\ell )^2 (\ell +2 \epsilon ) (\Delta +2 \Delta_\phi +\ell -2 \epsilon -2)^2}{32 (\Delta -\epsilon -1) (\Delta -\epsilon ) (\Delta +\ell -1) (\Delta +\ell +1) (\ell +\epsilon )}\;,\\
{}&\mathcal{U}=-\frac{(\Delta -1) \ell  (\Delta -2 \epsilon ) (-\Delta +\ell +2 \epsilon )^2 (-\Delta -2 \Delta_\phi +\ell +4 \epsilon +2)^2}{32 (\Delta -\epsilon -1) (\Delta -\epsilon ) (\ell +\epsilon ) (-\Delta +\ell +2 \epsilon -1) (-\Delta +\ell +2 \epsilon +1)}\;,\\
{}&\mathcal{W}=C^{(2)}_{\Delta_E,\ell_E}+\frac{1}{2}C^{(2)}_{\Delta,\ell}-2C^{(2)}_{\Delta_\phi,0}
\end{split}
\end{equation}
are obtained from $\mathfrak{R}$, $\mathfrak{S}$, $\mathfrak{T}$, $\mathfrak{U}$ and $\mathfrak{W}$ in (\ref{EOMrecuranydcoe}) by setting $\Delta_i=\Delta_\phi$. The coefficients $\mathcal{R}'$, $\mathcal{S}'$, $\mathcal{T}'$, $\mathcal{U}'$, $\mathcal{W}'$ are obtained from the unprimed coefficients by taking derivative with respect to $\Delta$.

Using this recursion relation and (\ref{EOMrecuranyd}) with $\Delta_i=\Delta_\phi$, we obtain the following relations for the crossed channel coefficients in (\ref{WexchangeintEW}). We have
\begin{equation}\label{recuranydEWOPEcoe}
\mathcal{R}_{n+1,\ell-1}\mathcal{C}_{n+1,\ell-1}+\mathcal{S}_{n,\ell+1}\mathcal{C}_{n,\ell+1}+\mathcal{T}_{n,\ell-1}\mathcal{C}_{n,\ell-1}
+\mathcal{U}_{n-1,\ell+1}\mathcal{C}_{n-1,\ell+1}+\mathcal{W}_{n,\ell}\mathcal{C}_{n,\ell}=\tilde{c}_{n,\ell}\;,
\end{equation}
and 
\begin{equation}\label{recuranydEWanom}
\begin{split}
{}&\mathcal{R}_{n+1,\ell-1}\mathcal{B}_{n+1,\ell-1}+\mathcal{S}_{n,\ell+1}\mathcal{B}_{n,\ell+1}+\mathcal{T}_{n,\ell-1}\mathcal{B}_{n,\ell-1}
+\mathcal{U}_{n-1,\ell+1}\mathcal{B}_{n-1,\ell+1}+\mathcal{W}_{n,\ell}\mathcal{B}_{n,\ell}\\
+{}&\mathcal{R}_{n+1,\ell-1}'\mathcal{C}_{n+1,\ell-1}+\mathcal{S}_{n,\ell+1}'\mathcal{C}_{n,\ell+1}+\mathcal{T}_{n,\ell-1}'\mathcal{C}_{n,\ell-1}
+\mathcal{U}_{n-1,\ell+1}'\mathcal{C}_{n-1,\ell+1}+\mathcal{W}_{n,\ell}'\mathcal{C}_{n,\ell}=\tilde{a}_{n,\ell}\;.
\end{split}
\end{equation}
Here $\mathcal{R}_{n,\ell}$, $\mathcal{R}_{n,\ell}'$ {\it etc.} are $\mathcal{R}$ and $\mathcal{R}'$ with $\Delta=2\Delta+2n+\ell$. The decomposition coefficients $\mathcal{B}_{n,\ell}$, $\mathcal{C}_{n,\ell}$ are related to the $B_{n,\ell}$, $C^{(t)}_{n,\ell}$ in (\ref{WexchangeintEW}) via
\begin{equation}
\mathcal{B}_{n,\ell}\mathcal{N}_{\epsilon,\ell}=B_{n,\ell}\;,\quad \mathcal{C}_{n,\ell}\mathcal{N}_{\epsilon,\ell}= C^{(t)}_{n,\ell}\;,
\end{equation}
and vanish when either $n$ or $\ell$ is negative. Finally $\tilde{c}_{n,\ell}$ and $\tilde{a}_{n,\ell}$ are sums of $c_{n,\ell}$ and $a_{n,\ell}$ in (\ref{WcontactinsEW}) for the contact diagrams that appear on the RHS of the equation of motion identity.

By setting the RHS' to zero in (\ref{recuranydEWOPEcoe}) and (\ref{recuranydEWanom}), we obtain the recursion relations for decomposing an equal weight t-channel conformal partial wave into the s-channel.  

\subsubsection*{Equal weight recursion relations for $d=1$}
We can repeat the analysis for $d=1$. The action of $\mathbf{EOM}^{(t)}$ on the 1d derivative block $\partial_\Delta g^{(s)}_{\Delta}(z)$ can be obtained from taking derivative with respect to $\Delta$ on (\ref{crosseomong}). This gives us
\begin{equation}\label{crosseomonderg}
\begin{split}
\mathbf{EOM}^{(t)}[\partial_\Delta g^{(s)}_{\Delta}(z)]={}&\mu\,\partial_\Delta g^{(s)}_{\Delta-1}(z)+\nu\,\partial_\Delta g^{(s)}_{\Delta}(z)+\rho\,\partial_\Delta g^{(s)}_{\Delta+1}(z)\\
{}&+\mu'\, g^{(s)}_{\Delta-1}(z)+\nu'\, g^{(s)}_{\Delta}(z)+\rho'\, g^{(s)}_{\Delta+1}(z)\;
\end{split}
\end{equation}
where 
\begin{equation}
\begin{split}
\mu={}&-(\Delta-2\Delta_\phi)^2\;,\\
\nu={}&\Delta_E(\Delta_E-1)+\frac{1}{2}\Delta(\Delta-1)-2\Delta_\phi(\Delta_\phi-1)\;,\\
\rho={}&-\frac{\Delta^2(-1+\Delta+2\Delta_\phi)^2}{4(2\Delta-1)(2\Delta+1)}\;.
\end{split}
\end{equation}
The coefficients $\mu'$, $\nu'$, $\rho'$ are obtained from the above expressions by taking derivative with respect to $\Delta$.

In the equal weight case, the s-channel conformal block decomposition of the Witten diagrams takes the following form
\begin{equation}
\mathcal{W}^{t,exchange}_{\Delta_E}(z)=\sum_{n=0}^\infty B_n g^{(s)}_{2\Delta_\phi+n}(z)+\sum_{n=0}^\infty C_n^{(t)} \partial g^{(s)}_{2\Delta_\phi+n}(z)\;,
\end{equation} 
\begin{equation}
\mathcal{D}_{\Delta_\phi\Delta_\phi\Delta_\phi\Delta_\phi}(z)=\sum_{n=0}^\infty a_n g^{(s)}_{2\Delta_\phi+2n}(z)+\sum_{n=0}^\infty c_n \partial g^{(s)}_{2\Delta_\phi+2n}(z)\;.
\end{equation} 
Using the above recursion relations in the t-channel equation of motion identity 
\begin{equation}
\mathbf{EOM}^{(t)}[\mathcal{W}^{t,exchange}_{\Delta_E}](z)=\mathcal{D}_{\Delta_\phi\Delta_\phi\Delta_\phi\Delta_\phi}(z)\;,
\end{equation}
we obtain the following recursion relations for the decomposition coefficients
\begin{equation}
\rho_{n-1}C^{(t)}_{n-1}+\nu_{n}C^{(t)}_{n}+\mu_{n+1}C^{(t)}_{n+1}=
\begin{cases}
c_{\frac{n}{2}}\;,\quad n\text{ even}\;,\\
0\;,\quad n\text{ odd}\;,
\end{cases}
\end{equation}
\begin{equation}
\rho_{n-1}B_{n-1}+\nu_{n}B_{n}+\mu_{n+1}B_{n+1}+\rho_{n-1}'C^{(t)}_{n-1}+\nu_{n}'C^{(t)}_{n}+\mu_{n+1}'C^{(t)}_{n+1}=
\begin{cases}
a_{\frac{n}{2}}\;,\quad n\text{ even}\;,\\
0\;,\quad n\text{ odd}\;.
\end{cases}
\end{equation}
Here we also impose $B_{-1}=C^{(t)}_{-1}=0$. The various coefficients $\mu_n$, $\mu_n'$ {\it etc}. are obtained from $\mu$ and $\mu'$ by setting $\Delta=2\Delta_\phi+n$. The coefficients $a_n$ and $c_n$ are obtained by using (\ref{Dfunctiondecomcoe}), and taking the equal weight limit and setting $d=1$
\begin{equation}
\begin{split}
a_n={}&\frac{\pi ^{1/2} (-1)^{-2 n} \Gamma (n+\Delta_\phi )^4 \Gamma \left(-\frac{1}{2}+n+2 \Delta_\phi \right)^2}{(n!)^2 \Gamma (\Delta_\phi )^4 \Gamma (2 (n+\Delta_\phi )) \Gamma \left(2 (n+\Delta_\phi )-\frac{1}{2}\right)}\\
{}&\times \left(-2 H_{n+\Delta_\phi -1}-H_{n+2 \Delta_\phi -\frac{3}{2}}+2 H_{4 n+4 \Delta_\phi -2}+H_n-\log (4)\right)\\
c_n={}&\frac{\pi ^{1/2} (-1)^{1-2 n} \Gamma (n+\Delta_\phi )^4 \Gamma \left(-\frac{1}{2}+n+2 \Delta_\phi \right)^2}{(n!)^2 \Gamma (\Delta_\phi )^4 \Gamma (2 (n+\Delta_\phi )) \Gamma \left(2 (n+\Delta_\phi )-\frac{1}{2}\right)}\;.
\end{split}
\end{equation}
The seed coefficients $B_0$, $C^{(t)}_0$ can be obtained from the unequal weight seed coefficients $B^{12}_0$, $B^{34}_0$ in Appendix \ref{appseed} by taking the equal weight limit. The explicit expression of $C^{(t)}_0$ is given by (\ref{Ct0}).

\section{Computing the Seed Coefficient for $\mathrm{CFT}_1$}\label{appseed}
In this appendix, we compute the seed coefficient for decomposing a t-channel scalar exchange Witten diagram in $AdS_2$ into the s-channel. The strategy is to use the method of \cite{DHoker:1999mqo} to integrate out the exchanged bulk field, and express the exchange diagram as infinitely many $D$-functions dressed with appropriate powers of $x_{ij}^2$. By taking the s-channel OPE limit, we can obtain the seed coefficient by resuming the contributions. 

We start with the three-point integral 
\begin{equation}
I^{t,exchange}(x_2,x_3;z_2)=\int_{AdS_{d+1}}\frac{d^{d+1}z_1}{z_{10}^{d+1}} {G}_{BB}^\Delta(z_2,z_1){G}_{B\partial}^{\Delta_2}(z_1,x_2){G}_{B\partial}^{\Delta_3}(z_1,x_3)\;.
\end{equation}
The t-channel scalar exchange diagram is obtained by further performing a $z_2$ integral
\begin{equation}\label{z2integral}
W^{t,exchange}(x_i)=\int_{AdS_{d+1}}\frac{d^{d+1}z_2}{z_{20}^{d+1}}I^{t,exchange}(x_2,x_3;z_2)
{G}_{B\partial}^{\Delta_1}(z_2,x_1){G}_{B\partial}^{\Delta_4}(z_2,x_4)\;.
\end{equation}
At this stage, we will keep the boundary spacetime dimension $d$ general, and will only set $d=1$ at the very end.
Following \cite{DHoker:1999mqo}, we perform a translation such that
\begin{equation}
x_2\to 0\;,\;\;\;x_3\to x_{32}\equiv x_3-x_2\;.
\end{equation}
This is followed by a conformal inversion,
\begin{equation}
x_{23}'=\frac{x_{23}}{(x_{23})^2}\;,\;\;\; z_1'=\frac{z_1}{z_1^2}\;,\;\;\; z_2'=\frac{z_2}{z_2^2}.
\end{equation}
After making these transformations the three-point integral becomes,
\begin{equation}\label{z1integralresult}
I^{t,exchange}(x_2,x_3;z_2)=(x_{23})^{-2\Delta_3}J(z_2'-x_{23}')
\end{equation}
where we have defined
\begin{equation}
J(z_2)=\int\frac{d^{d+1}z_1}{z_{10}^{d+1}}\;\Pi^{\Delta}(u)\;z_{10}^{\Delta_2}\left(\frac{z_{10}}{z_1^2}\right)^{\Delta_3}\;.
\end{equation}
We can rewrite $J(z_2)$ in the following form
\begin{equation}
J(z_2)=z_{20}^{\Delta_2-\Delta_3} f(t)\;,
\end{equation}
using its scaling behavior under $z_2\to\lambda z_2$ and Poincar\'e symmetry. Here for convenience we defined
\begin{equation}
t=\frac{z_{20}^2}{z_2^2}\;,
\end{equation}
and it takes values in $[0,1]$ in the physical regime. On the other hand, because the bulk-to-bulk propagator satisfies the equation of motion, we can derive the following differential equation for $f(t)$
\begin{equation}\label{diffeqnfoft}
4t^2(t-1)f''+4t[(\Delta_2-\Delta_3+1)t-\Delta_2+\Delta_3+\frac{d}{2}-1]f'+[(\Delta_2-\Delta_3)(d-\Delta_2+\Delta_3)+M_E^2]f=t^{\Delta_3}
\end{equation}
This differential equation is further supplemented by two boundary conditions: 
\begin{enumerate}
\item[1)] From the OPE limit, we know that $f(t)$ should behave like
\begin{equation}
f(t)\sim t^{\frac{\Delta_E-\Delta_2+\Delta_3}{2}}\;,\quad t\to 0\;.
\end{equation}
\item[2)] 
From definition of the integral, $f(t)$ has to be smooth at $t=1$  (see \cite{DHoker:1998ecp}). 
\end{enumerate}
We find that the solution to $f(t)$ with appropriate boundary conditions is given by
\begin{equation}\small\label{solutionfoft}
f(t)=C_s t^{\Delta_3} {}_3F_2\left(1,\Delta_2,\Delta_3;-\frac{\Delta_E}{2}+\frac{\Delta_2}{2}+\frac{\Delta_3}{2}+1,-\frac{d}{2}+\frac{\Delta_E }{2}+\frac{\Delta_2}{2}+\frac{\Delta_3}{2}+1;t\right)+C_hf_h(t)
\end{equation}
where
\begin{equation}\small
\begin{split}
C_s={}&-\frac{1 \, }{(-\Delta_E +\Delta_2+\Delta_3) (-d+\Delta_E +\Delta_2+\Delta_3)}\;,\\
C_h={}&\frac{\Gamma \left(\frac{\Delta_E +\Delta_2-\Delta_3}{2}\right) \Gamma \left(\frac{\Delta_E -\Delta_2+\Delta_3}{2}\right) \Gamma \left(\frac{-\Delta_E +\Delta_2+\Delta_3}{2}\right) \Gamma \left(\frac{-d+\Delta_E +\Delta_2+\Delta_3}{2}\right)}{4 \Gamma (\Delta_2) \Gamma (\Delta_3) \Gamma \left(-\frac{d}{2}+\Delta +1\right)}\;,
\end{split}
\end{equation}
and
\begin{equation}\small
f_h(t)=t^{\frac{1}{2} (\Delta_E -\Delta_2+\Delta_3)} \, _2F_1\left(\frac{1}{2} (\Delta_E -\Delta_2+\Delta_3),\frac{1}{2} (\Delta_E +\Delta_2-\Delta_3);-\frac{d}{2}+\Delta_E +1;t\right)
\end{equation}
is a homogenous solution to (\ref{diffeqnfoft}). In particular, when the external conformal dimension satisfies the relation $\Delta_2+\Delta_3-\Delta_E=2m$ where $m\in\mathbb{Z}^+$, the above solution truncates to a polynomial $f^{poly}(t)$ \cite{DHoker:1999mqo}
\begin{equation}
f^{poly}(t)=\sum_{k=k_{\min}}^{k_{\max}}a_kt^k
\end{equation}
where 
\begin{equation}
\begin{split}
k_{\min}={}&\frac{\Delta_E-\Delta_2+\Delta_3}{2}\;,\\
k_{\max}={}&\Delta_3-1\;,\\
a_{k_{\max}}={}&\frac{1}{4(\Delta_2-1)(\Delta_3-1)}\;,\\
a_{k-1}={}&\frac{(k-\frac{\Delta_E}{2}+\frac{\Delta_2-\Delta_3}{2})(k-\frac{d}{2}+\frac{\Delta_E}{2}+\frac{\Delta_2-\Delta_3}{2})}{(k-1)(k-1-\Delta_2+\Delta_3)}a_k\;.
\end{split}
\end{equation}
We can also write the solution (\ref{solutionfoft}) as a sum of two infinite series
\begin{equation}\label{fsolseries}
f(t)=t^{\Delta_3}\sum_{i=0}^\infty P_i\,t^i+t^{\frac{\Delta_E-\Delta_2+\Delta_3}{2}}\sum_{i=0}^\infty Q_i\,t^i
\end{equation}
where
\begin{equation}\small
P_i=\frac{(\Delta_2)_i (\Delta_3)_i}{(\Delta_E -\Delta_2-\Delta_3) (-d+\Delta_E +\Delta_2+\Delta_3) \left(\frac{-\Delta_E +\Delta_2+\Delta_3+2}{2}\right)_i \left(\frac{-d+\Delta_E +\Delta_2+\Delta_3+2}{2}\right)_i}\;,
\end{equation}
and
\begin{equation}\small
\begin{split}
Q_i={}&\frac{(-1)^i \Gamma \left(\frac{d-2 i-2\Delta_E}{2}\right)\sin \left(\frac{\pi  (d-2 \Delta_E )}{2} \right)\Gamma \left(\frac{-d+\Delta_E +\Delta_2+\Delta_3}{2}\right)}{4 \pi  \Gamma (i+1)\Gamma (\Delta_2) \Gamma (\Delta_3)}\\
{}&\times \frac{\Gamma \left(\frac{\Delta_E -\Delta_2+\Delta_3}{2}\right) \Gamma \left(\frac{\Delta_E +\Delta_2-\Delta_3}{2}\right) \Gamma \left(\frac{-\Delta_E +\Delta_2+\Delta_3}{2}\right)\Gamma \left(\frac{-\Delta_E +\Delta_2-\Delta_3+2}{2}\right)  \Gamma \left(\frac{-\Delta_E -\Delta_2+\Delta_3+2}{2}\right) }{\Gamma \left(\frac{-\Delta_E +\Delta_2-\Delta_3-2 i+2}{2}\right)\Gamma \left(\frac{-\Delta_E -\Delta_2+\Delta_3-2 i+2}{2}\right)}\;.
\end{split}
\end{equation}
After obtaining the solution for $f(t)$, we can undo the transformations to get (\ref{z1integralresult}). The upshot is that every power $t^a$ becomes a contact vertex
\begin{equation}
(x_2-x_3)^{2(a-\Delta_3)}G_{B\partial}^{a+\Delta_2-\Delta_3}(x_2,z_2)\;G_{B\partial}^{a}(x_3,z_2)\;.
\end{equation}
Using this replacement in (\ref{fsolseries}) and perform the $z_2$ integral, we obtain the following result for (\ref{z2integral})
\begin{equation}\small\label{WtexchangeasDfun}
W^{t,exchange}=\sum_{i=0}^\infty (x_{23}^2)^{i} P_i\, D_{\Delta_1,\Delta_2+i,\Delta_3+i,\Delta_4}+\sum_{i=0}^\infty (x_{23}^2)^{\frac{\Delta_E-\Delta_2-\Delta_3+2i}{2}} Q_i\, D_{\Delta_1,\frac{\Delta_E+\Delta_2-\Delta_3}{2}+i,\frac{\Delta_E-\Delta_2+\Delta_3}{2}+i,\Delta_4}\;.
\end{equation}
As a consistency check, let us extract several t-channel OPE coefficients from this expression. Taking the t-channel OPE limit $x_2\to x_3$, each $D$-function becomes a three-point function (\ref{D3function})
\begin{equation}
D_{\Delta_1\Delta_2\Delta_3\Delta_4}\xrightarrow{x_{23}^2\to 0}D_{\Delta_1,\Delta_2+\Delta_3,\Delta_4}\;.
\end{equation}
It is easy to see that (\ref{WtexchangeasDfun}) becomes
\begin{equation}\label{tleadingWtexchange}
W^{t,exchange}=P_0D_{\Delta_1,\Delta_2+\Delta_3,\Delta_4}\left[1+\mathcal{O}\left(\frac{x_{23}^2}{\Lambda}\right)\right]+Q_0(x_{23}^2)^{\frac{\Delta_E-\Delta_2-\Delta_3}{2}}D_{\Delta_1,\Delta_E,\Delta_4}\left[1+\mathcal{O}\left(\frac{x_{23}^2}{\Lambda}\right)\right]
\end{equation}
in this limit, where $\Lambda$ is some dimensionful quantity ({\it e.g.}, $x_{12}^2$ and $x_{14}^2$) that makes the ratio dimensionless. The first term corresponds to the contribution from the leading double-trace operator $\Delta_2+\Delta_3$, for which the coefficient
\begin{equation}
P_0\, a_{\Delta_1,\Delta_2+\Delta_3,\Delta_4}
\end{equation}
 reproduces the OPE coefficient $A^{12}_{0,0}$ upon switching $\Delta_1\leftrightarrow \Delta_3$ in (\ref{A12andA34}). The second term in (\ref{tleadingWtexchange}) corresponds to leading contribution from the single-trace operator exchange. The coefficient 
\begin{equation}
Q_0\, a_{\Delta_1,\Delta_E,\Delta_4}
\end{equation}
correctly reproduces the single-trace OPE coefficient $A$ in (\ref{Wexchangeins}) after interchanging $\Delta_1$ and $\Delta_3$.

Now let us extract the seed coefficient $B^{12}_{0}$ in (\ref{texchange1dB12B34}) for an 1d scalar exchange Witten diagram. We set $d=1$ in (\ref{WtexchangeasDfun}) and take $x_{12}^2\to 0$. We can further assume $\Delta_1+\Delta_2<\Delta_3+\Delta_4$ so that the $x_{12}^2\to 0$ is dominated by the exchange of the double-trace operator $:O_1O_2:$. Then
\begin{equation}
\begin{split}
W^{t,exchange}\xrightarrow{x_{12}^2\to 0}{}&\frac{\sum_iP_i\,a_{\Delta_1+\Delta_2+i,\Delta_3+i,\Delta_4}+\sum_iQ_ia_{\Delta_1+\frac{\Delta_E+\Delta_2-\Delta_3}{2}+i,\frac{\Delta_E-\Delta_2+\Delta_3}{2}+i,\Delta_4}}{|x_{13}|^{\Delta_1+\Delta_2+\Delta_3-\Delta_4}|x_{14}|^{\Delta_1+\Delta_2+\Delta_4-\Delta_3}|x_{34}|^{\Delta_3+\Delta4+\Delta_1-\Delta_2}}\\
{}&=B^{12}_0 g^{(s)}_{\Delta_1+\Delta_2}(x_i)\bigg|_{x_{12}^2=0}\\
{}&=\frac{B^{12}_0}{|x_{13}|^{\Delta_1+\Delta_2+\Delta_3-\Delta_4}|x_{14}|^{\Delta_1+\Delta_2+\Delta_4-\Delta_3}|x_{34}|^{\Delta_3+\Delta4+\Delta_1-\Delta_2}}\;,
\end{split}
\end{equation}
and we get
\begin{equation}\label{A120}
\begin{split}
B^{12}_0={}&X{}_4F_3\left(\left.\begin{array}{c}1,\Delta_2,\frac{\Delta_1}{2}+\frac{\Delta_2}{2}+\frac{\Delta_3}{2}-\frac{\Delta_4}{2},\frac{\Delta_1}{2}+\frac{\Delta_2}{2}+\frac{\Delta_3}{2}+\frac{\Delta_4}{2}-\frac{1}{2} \\\Delta_1+\Delta_2,-\frac{\Delta_E }{2}+\frac{\Delta_2}{2}+\frac{\Delta_3}{2}+1,\frac{\Delta_E }{2}+\frac{\Delta_2}{2}+\frac{\Delta_3}{2}+\frac{1}{2}\end{array}\right.;1\right)\\
{}&+Y{}_3F_2\left(\left.\begin{array}{c} \frac{\Delta_E }{2}+\frac{\Delta_2}{2}-\frac{\Delta_3}{2},\frac{\Delta_E }{2}+\frac{\Delta_1}{2}-\frac{\Delta_4}{2},\frac{\Delta_E }{2}+\frac{\Delta_1}{2}+\frac{\Delta_4}{2}-\frac{1}{2} \\ \Delta_E +\frac{1}{2},\frac{\Delta_E }{2}+\Delta_1+\frac{\Delta_2}{2}-\frac{\Delta_3}{2}\end{array}\right.;1\right)
\end{split}
\end{equation}
where
\begin{equation}
X=\frac{\sqrt{\pi } \Gamma \left(\frac{\Delta_1+\Delta_2+\Delta_3-\Delta_4}{2}\right) \Gamma \left(\frac{\Delta_1+\Delta_2-\Delta_3+\Delta_4}{2}\right) \Gamma \left(\frac{-\Delta_1-\Delta_2+\Delta_3+\Delta_4}{2}\right) \Gamma \left(\frac{\Delta_1+\Delta_2+\Delta_3+\Delta_4-1}{2}\right)}{2 \Gamma (\Delta_3) \Gamma (\Delta_4) (\Delta -\Delta_2-\Delta_3) (\Delta_E +\Delta_2+\Delta_3-1) \Gamma (\Delta_1+\Delta_2)}\;,
\end{equation}
\begin{equation}
\begin{split}
Y={}&\frac{\sqrt{\pi } \Gamma \left(\frac{\Delta_E +\Delta_1-\Delta_4}{2}\right) \Gamma \left(\frac{\Delta_E +\Delta_1+\Delta_4-1}{2}\right) \Gamma \left(\frac{\Delta_E +\Delta_2-\Delta_3}{2}\right) \Gamma \left(\frac{-\Delta_E +\Delta_2+\Delta_3}{2} \right) \Gamma \left(\frac{\Delta_E +\Delta_2+\Delta_3-1}{2} \right) }{8 \Gamma \left(\Delta_E +\frac{1}{2}\right) \Gamma (\Delta_2) \Gamma (\Delta_3) \Gamma (\Delta_4)}\\
{}&\times \frac{\Gamma \left(\frac{\Delta_1+\Delta_2-\Delta_3+\Delta_4}{2}\right) \Gamma \left(\frac{-\Delta_1-\Delta_2+\Delta_3+\Delta_4}{2}\right)}{ \Gamma \left(\frac{\Delta_E +2 \Delta_1+\Delta_2-\Delta_3}{2}\right)}\;.
\end{split}
\end{equation}
The seed coefficient $B^{34}_0$ can be worked out similarly, and the result is just $B^{12}_0$ with the replacement $\Delta_1\leftrightarrow\Delta_3$, $\Delta_2\leftrightarrow\Delta_4$.

Taking the equal weight limit $\Delta_i\to\Delta_\phi$ for
\begin{equation}
B^{12}_0g^{(s)}_{\Delta_1+\Delta_2}+B^{34}_0g^{(s)}_{\Delta_3+\Delta_4}\;,
\end{equation}
we obtain the seed coefficients $B_0$ and $C^{(t)}_0$
\begin{equation}
B_0g^{(s)}_{2\Delta_\phi}+C^{(t)}_0\partial g^{(s)}_{2\Delta_\phi}\;.
\end{equation}
We record here the simpler coefficient $C^{(t)}_0$ for reader's convenience
\begin{equation}\label{Ct0}
\begin{split}
C^{(t)}_0={}&-\frac{\sqrt{\pi } \Gamma \left(2 \Delta_\phi -\frac{1}{2}\right)}{(\Delta_E -2 \Delta_\phi ) (\Delta_E +2 \Delta_\phi -1) \Gamma (2 \Delta_\phi )}{}_4F_3\left(\left.\begin{array}{c}1,\Delta_\phi,\Delta_\phi,-\frac{1}{2}+2\Delta_\phi \\2\Delta_\phi,1-\frac{\Delta_E}{2}+\Delta_\phi,\frac{1}{2}+\frac{\Delta_E}{2}+\Delta_\phi\end{array}\right.;1\right)\\
{}&-\frac{\sqrt{\pi } \Gamma \left(\frac{\Delta_E }{2}\right)^2 \Gamma \left(\frac{\Delta_E -1}{2}+\Delta_\phi \right)^2 \Gamma \left(\Delta_\phi -\frac{\Delta_E }{2}\right)}{4 \Gamma \left(\Delta_E +\frac{1}{2}\right) \Gamma (\Delta_\phi )^2 \Gamma \left(\frac{\Delta_E}{2}+\Delta_\phi \right)} {}_3F_2\left(\left.\begin{array}{c}\frac{\Delta_E}{2},\frac{\Delta_E}{2},-\frac{1}{2}+\frac{\Delta_E}{2}+\Delta_\phi \\\frac{1}{2}+\Delta_E,\frac{\Delta_E}{2}+\Delta_\phi\end{array}\right.;1\right)\;,
\end{split}
\end{equation}
while we refrain from printing out the explicit expression of the lengthier $B_0$. Obtaining $B_0$ from the above limit is a simple exercise.

\bibliography{Wittendiagrams} 
\bibliographystyle{utphys}

\end{document}